\documentclass[journal=apchd5,manuscript=article]{achemso}
\usepackage{upgreek}
\usepackage[version=3]{mhchem} 
\usepackage[numbers]{natbib}

\makeatletter

\author{Mahdad Mansouree}

\author{Andrew McClung}

\author{Sarath Samudrala}

\author{Amir Arbabi}

\email{arbabi@umass.edu}

\affiliation[umass]
{Department of Electrical and Computer Engineering,
University of Massachusetts Amherst, 151 Holdsworth Way, Amherst, MA 01003, USA}

\title[An \textsf{achemso} demo]
  {Large-scale parameterized metasurface design using adjoint optimization}

\abbreviations{IR,NMR,UV}

\begin{document}

\begin{abstract}
Optical metasurfaces are planar arrangements of subwavelength meta-atoms that implement a wide range of transformations on incident light.
The design of efficient metasurfaces requires that the responses of and interactions among meta-atoms are accurately modeled.
Conventionally, each meta-atom’s response is approximated by that of a meta-atom located in a periodic array.
Although this approximation is accurate for metastructures with slowly varying meta-atoms, it does not accurately model the complex interactions among meta-atoms in more rapidly varying metasurfaces.
Optimization-based design techniques that rely on full-wave simulations mitigate this problem but thus far have been mostly applied to topology optimization of small metasurfaces. 
Here, we describe an adjoint-optimization-based design technique that uses parameterized meta-atoms. Our technique has a lower computational cost than topology optimization approaches, enabling the design of large-scale metasurfaces that can be readily fabricated.
As proof of concept, we present the design and experimental demonstration of high numerical aperture metalenses with significantly higher efficiencies than their conventionally-designed counterparts. 
\end{abstract}

\section*{Introduction}

Optical metasurfaces are arrangements of subwavelength meta-atoms that scatter optical waves and generate desirable wavefront, amplitude, and polarization distributions \cite{Kamali2018}.
Metasurface-based designs of numerous optical components have been demonstrated, including lenses \cite{chen1996, Arbabi2015pol, Chen2012}, blazed gratings, \cite{Lalanne1998} and holograms \cite{Huang2013,Zhang2016,Zheng2015,Larouche2012}$\!$. Their planar form factor and the potential for low-cost manufacture have spurred the recent development of complex optical systems made of multiple metasurfaces, or metasystems, such as miniaturized cameras \cite{Arbabi2016camera}$\!$, spectrometers \cite{Faraji-Dana} and hyper-spectral imaging systems \cite{faraji2019hyperspectral}$\!$.
However, cascading multiple metasurfaces quickly increases a system’s optical loss and high-performance metasystems require high-efficiency metasurface components.

Currently, most metasurfaces are designed using a unit-cell-based approach in which each meta-atom is simulated as a part of a periodic array \cite{Arbabi2015a, chen1996, Lalanne1998, Yu}$\!$.
This approach is computationally inexpensive, readily scalable to arbitrarily large structures, and produces designs that can be fabricated easily.
However, two implicit assumptions in the unit-cell approach can lead to inefficient metasurface designs: First, the response of a meta-atom with dissimilar neighbors differs from the response of the same meta-atom in an array with identical elements. 
The ‘local periodicity’ approximation breaks down in structures with rapidly varying meta-atoms, such as high numerical aperture (NA) lenses. 
This approximation also has reduced accuracy in structures comprising meta-atoms with lower refractive index, in which the response of a meta-atom is more strongly affected by variations of its neighbors \cite{Bayati_role, Yang_analysis}.
Second, the response map used in unit-cell-based methods records only the normally transmitted response for normally incident light. 
In an actual metasurface, the transmission angle varies, and hence the true response would deviate from the response map \cite{Mahsa_ACS}. 
These assumptions can lead to significant mismatches between expected and actual responses of a meta-atom in a metasurface. The effect of such mismatch can be seen in reduction of the efficiency of the device, however, it is hard to exactly distinguish the contribution of each assumption without analytical models.

In the absence of simple and accurate models that capture the individual and collective behaviors of meta-atoms, optimization-based design (i.e., inverse design) is a practical alternative. Optical structures have been designed using a variety of optimization-based approaches. A comprehensive review of these methods is presented in ref. \cite{molesky2018inverse}$\!$. Heuristic optimization algorithms based on random explorations of the optimization space (e.g., particle swarm optimization or genetic algorithms) have been used to design diffraction grating filters \cite{shokooh2007particleswarm}, polarization beam splitters \cite{shen2015particleswarm} and other small structures. Heuristic optimization is well-suited to problems with a small number of degrees of freedom but inefficient for structures with larger design spaces \cite{sigmund2011usefulness}. 

In most problems, the gradient of the design objective, necessary for gradient-based algorithms, can be determined using an adjoint technique. Adjoint-based algorithms are suitable for optimization spaces with high dimensionality\cite{Jensen2011}$\!$, and have been used to design high-performance optical devices, including grating couplers \cite{Niederberger:14}$\!$, polarization beam splitters\cite{Lalau-Keraly2013}$\!$, photonic crystals\cite{burger2004inverse}$\!$, metagratings \cite{Lalau-Keraly2013}$\!$ and metasurfaces\cite{sell2017, Phan2018, sell2017}$\!$. Adjoint-based optimization is frequently applied to topology optimization problems \cite{tsuji2006design, Niederberger:14,Lalau-Keraly2013, burger2004inverse, sell2017, Phan2018, lin2019topology, Jensen2011, yang2017, lin2019topology, Chung2020highNA, phan2019high, bayati2020inverse}$\!$, in which a structure or its meta-atoms are defined by a large number of pixels or sets of curvilinear patterns \cite{Jensen2011, yang2017}$\!$.
This can lead to efficient metasurface designs, but because even deeply subwavelength changes in meta-atom geometries can significantly alter the scattering response of a design (see Supplementary Note 1 and Fig.~S1), the meta-atom geometries should be accurately approximated during simulations.
The accurate representation of meta-atoms with arbitrary shapes require high-resolution meshings, practically limiting the structures to 2D  \cite{bayati2020inverse, phan2019high, Chung2020highNA, lin2019overlapping, Perez-Arancibia:18}$\!$ or small 3D \cite{Chung2020highNA} designs.
As a result, the technique has been mostly used for optimizing periodic structures such as gratings and cylindrical structures \cite{Sigmund2009tolerant, Wang:11robust, sell2017, lin2019topology}$\!$. 


To address this limitation, topology optimization recently has been combined with a local periodicity approximation \cite{phan2019high, lin2019topology, lin2019overlapping, Perez-Arancibia:18}.
In this approach, topology optimization is done on small subdomain of the device whose response within the larger structure is approximated by periodic boundary conditions. These subdomains are subsequently stitched together to form the large-scale structure. 
This approach enables the optimization of large devices; however, subdomains with periodic boundaries do not accurately model the local response in high-NA metalenses or other rapidly-varying structures, limiting the performance of designs arrived at by this approach.

\begin{figure}
    \centering
    \includegraphics{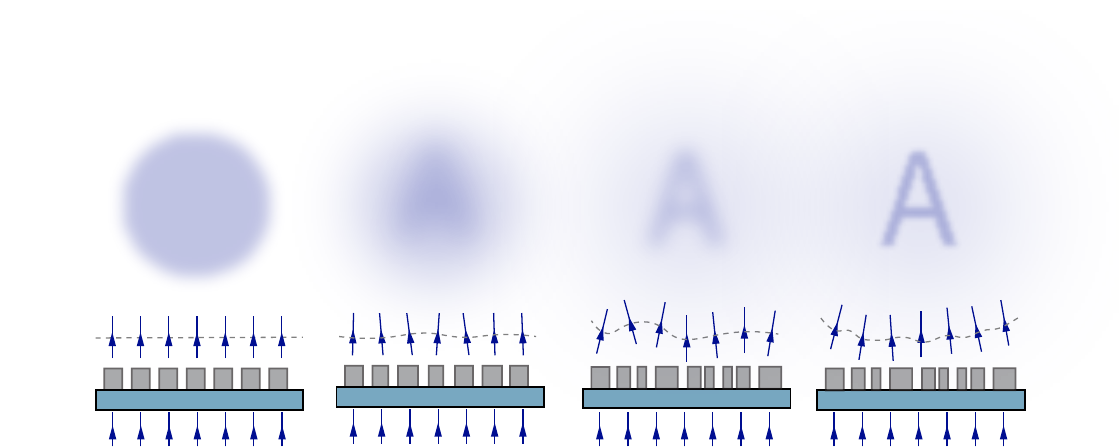}
    \caption{Illustration of the parameterized metasurface design process using adjoint optimization. As the structure is updated by the optimization method, the desired output fields start to form.}
    \label{fig:opto}
\end{figure}


Instead of designing free-form structures, here we propose and demonstrate an adjoint optimization method based on parameterized rectangular meta-atoms (Fig.~\ref{fig:opto}). 
Parameterized meta-atoms lack the fine features typical of topology-optimized structures , enabling simulations to converge at relatively low resolution and thus very large metasurfaces to be designed.
Confining the design space to simple shapes (e.g. rectangular meta-atoms) also reduces the cost of simulation preprocessing steps like subpixel smoothing \cite{farjadpour2006improving, Oskooi2010} (Supplementary Note 2 and Supplementary Fig.~S2). 
More importantly, limiting the optimization to this specific subspace of structures (i.e., rectangular meta-atoms)
removes a large number of potential local optima traps without significantly affecting device performance and produces designs that conform to a well-established metasurface platform that can be easily fabricated.
Our method also relies on a field interpolation technique for $\mathbf{E}_\parallel$ and $D_\bot$ (see methods) and an efficient time to frequency domain conversion technique to reduce the computational cost of simulating large structures.
Our method relies on full-wave, finite difference time domain (FDTD) simulations of the entire structure, and iteratively approaches an optimal design via gradient ascent.
A similar parameterized approach based on Mie theory was recently proposed by Zhan et. al. \cite{Zhan:18, zhan2019controlling}$\!$. However, the approach is limited to spherical meta-atoms, which are challenging to fabricate, and does not account for the substrate’s effect. 
The adjoint optimization technique does not rely on the two implicit assumptions used in the unit-cell approach, and thereby achieves higher performing designs. First, the variation in coupling among the meta-atoms caused by the rapid variation of their dimensions is accounted for. Second, no assumption is made about angular dependence of the meta-atom scattering response (i.e., its element factor).
In the following, we describe our method, and, as proof of concept, use it to design and fabricate two metalenses with 50 $\upmu$m diameter.
The focusing efficiencies of metalenses designed using this method show experimental improvements of 24\% and 13\% for NAs of 0.78 and 0.95 over counterparts designed by the conventional periodic unit-cell method.

\section*{Results}
\subsection*{Parameterized adjoint design method}
\begin{figure*}
	\centerline{\includegraphics{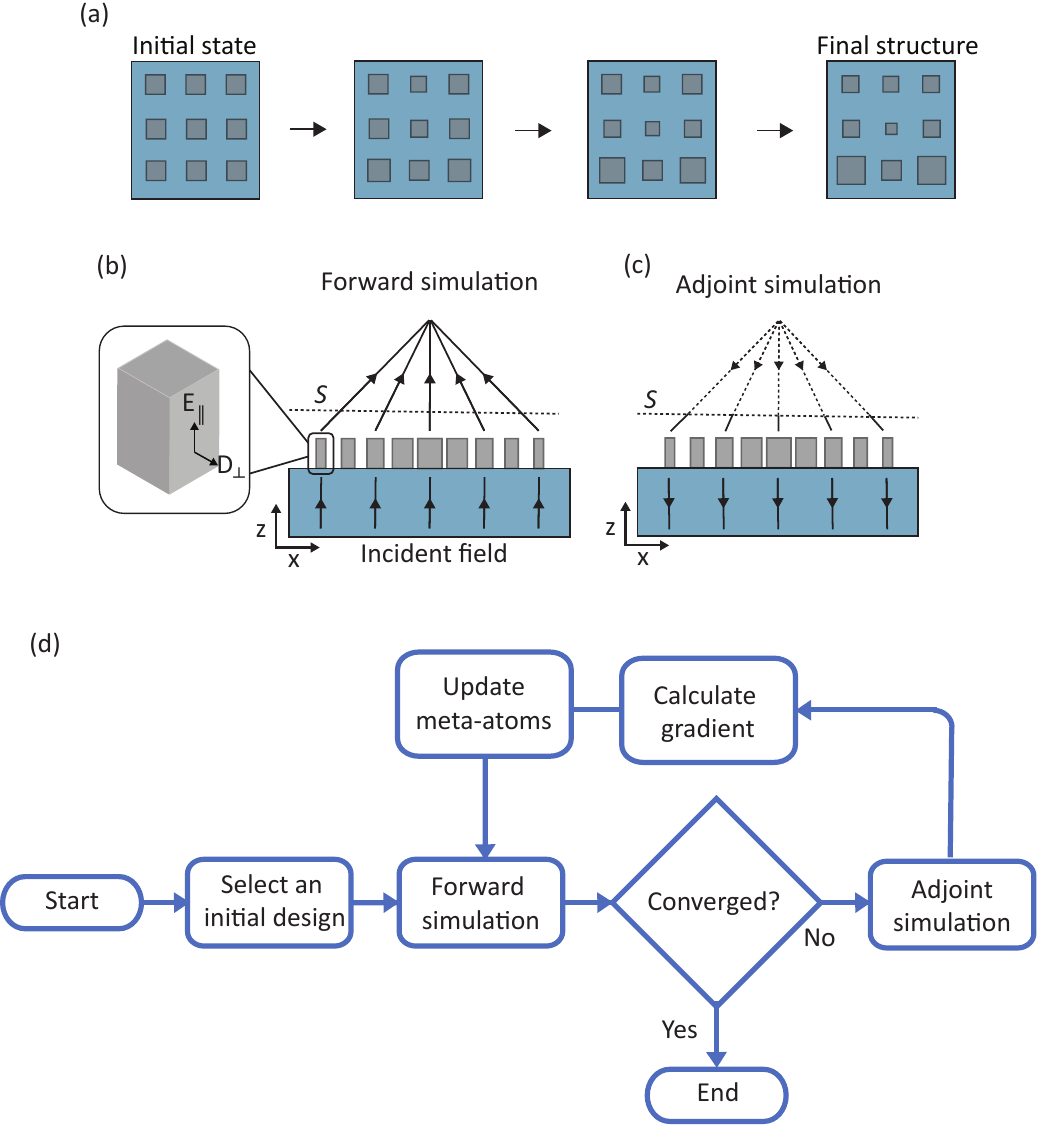}}
	\caption{Parameterized adjoint optimization. (a) Parameterized optimization: meta-atom dimensions are updated but constrained to a simple shape. (b) Representation of the forward problem. (c) Representation of the adjoint problem. (d) Flow diagram showing steps in the optimization. }
	\label{fig:flow_chart}
\end{figure*}

We first describe the metastructure design using the parameterized adjoint optimization method. The design process involves finding a set of meta-atom parameters that generate the desired transformation efficiently. As shown in Fig.~\ref{fig:flow_chart}a, to find the optimal design, we optimize the structure iteratively from a trivial initial design (e.g., a uniform array in which all meta-atoms are identical). In each iteration, the gradient of the objective function with respect to all the design parameters is calculated using only two simulations as conceptually shown in Fig.~\ref{fig:flow_chart}b-c.  Based on the computed gradient, each meta-atom is updated, generating a new design that is one step closer to the optimal design.  This cycle continues until all the parameters converge to their final values (Fig.~\ref{fig:flow_chart}d). 

 
 The goal of metastructure design is to transform an incident electric field $\mathbf{E}^\text{i}$ into a desired transmitted or reflected field distribution $\mathbf{E}^{\text{d}}$. This is schematically illustrated in Fig.~\ref{fig:flow_chart}b, which shows a metasurface transforming a normally incident field into a converging transmitted field. An arbitrary desired output can be achieved by specifying an appropriate transmitted field distribution on a plane $S$ above the metasurface. The metasurface we consider consists of an arrangement of dissimilar meta-atoms positioned on a periodic lattice. Each meta-atom’s shape is described by one or more parameters that are variables in the design process. Thus a design can be expressed as a vector ${\mathbf{p}}$ containing all the design parameters. 
In our proposed method, the design is cast as an optimization problem of maximizing the fraction of the output field in the desired field distribution. Specifically, an optimal design maximizes $I=\left|F\right|^2$, where

\begin{equation} \label{eq:1}
F(\mathbf{p})=\int_{S}^{\ }{\mathbf{E}^{\text{d}\ast}\cdot\mathbf{E}^\text{f}\left(\mathbf{p}\right)\ \text{d}A.}
\end{equation}
Here $\mathbf{E}^\text{d}$ is the desired field in the frequency domain on the plane $S$, $\mathbf{E}^\text{f}\left(\mathbf{p}\right)$ is the field realized by a design defined by $\mathbf{p}$ in the forward simulation excited by $\mathbf{E}^\text{i}$ (Fig.~\ref{fig:flow_chart}b), and * represents the complex conjugate operation. $F$ is the  complex-valued projection of $\mathbf{E}^\text{f}$ on $\mathbf{E}^\text{d}$.

Optimization starts from an initial design $\mathbf{p}^{(0)}$ and is updated iteratively $(\mathbf{p}^{(1)},\ \mathbf{p}^{(2)},\ \mathbf{\ldots})$ via gradient ascent. This process is illustrated in Fig.~\ref{fig:flow_chart}a: after each iteration, $\mathbf{p}$ approaches its locally optimal value and the performance of the metasurface improves. The gradient $\nabla_\mathbf{p}I$ is used to determine how $\mathbf{p}$ changes in the next step  and can be computed using an additional simulation called the adjoint simulation. The adjoint simulation uses the same design $\mathbf{p}$ as the forward simulation, but the structure is instead excited by a surface current density $\mathbf{J}_\text{s}^\text{a}$ $\equiv\mathbf{E}^{\text{d}\ast}$ that is placed on the plane $S$ which generates a backward propagating wave (see Fig.~\ref{fig:flow_chart}c). The electric field in the adjoint simulation is denoted $\mathbf{E}^\text{a}(\mathbf{p})$.

In general, the variation of $F$ with respect to small changes in the boundaries of meta-atoms can be found using the functional derivative of $F$. An expression for the functional derivative of $F$ based on symmetries of the Green’s tensor can be found in Ref. \cite{Miller2012a}$\!$. Here, we consider the special case of rectangular meta-atoms with square cross-sections (inset of Fig.~\ref{fig:flow_chart}b). For such meta-atoms, $\mathbf{p}=\left(w_1,\ w_2,..,w_N\right)$, where $w_i$ represents the width of $i^{th}$ meta-atom. Based on the Lorentz reciprocity theorem \cite{Harrington2001}$\!$, we show in Supporting Note 3 that the partial derivative of $F$ with respect to $w_i$ is given by

\begin{equation}\label{eq:2}
\frac{\partial F}{\partial w_i} = \frac{1}{2}~j\omega\left(n_\text{m}^2-n_\text{c}^2\right)\int_{\partial\Omega_i}^{\ }{\left(\mathbf{E}_\parallel^\text{f}\cdot\mathbf{E}_\parallel^\text{a}+\frac{1}{n_\text{m}^2n_\text{c}^2}D_\bot^\text{f}D_\bot^\text{a}\ \right)\mathrm{d}A,} 
\end{equation}
where $\omega$ is the angular frequency of the excitation, $n_\mathrm{m}$ and $n_\mathrm{c}$ are the refractive indices of meta-atom and cladding materials, ${\partial \Omega}_i$ represents the four side surfaces of the $i$th meta-atom, $\mathbf{E}_\parallel^\text{f}$ and $D_\bot^\text{f}$ are the tangential components of the electric field and normal component of the displacement field obtained in the forward problem, and $\mathbf{E}_\parallel^\text{a}$ and $D_\bot^\text{a}$ are the corresponding fields in the adjoint problem.  The gradient of the objective function necessary to determine the next design is given by $\nabla_\mathbf{p}I=2 \operatorname{Re} \left\{F^\ast\nabla_\mathbf{p}F\right\}$, where $\mathrm{Re}\left\{\cdot\right\}$ represents the real part of a complex number. 
Both forward and adjoint simulations are performed using full-wave simulations of the entire metasurface.
Although this is computationally more expensive than techniques that employ local periodicity approximations \cite{phan2019high, lin2019topology, lin2019overlapping, Perez-Arancibia:18}, it allows the gradient to be calculated more accurately.
The flow diagrams in Fig.~\ref{fig:flow_chart}d and Fig.~S3 summarize the optimization procedure.

\subsection*{Metalens design example}

\begin{figure*}
	\centerline{\includegraphics{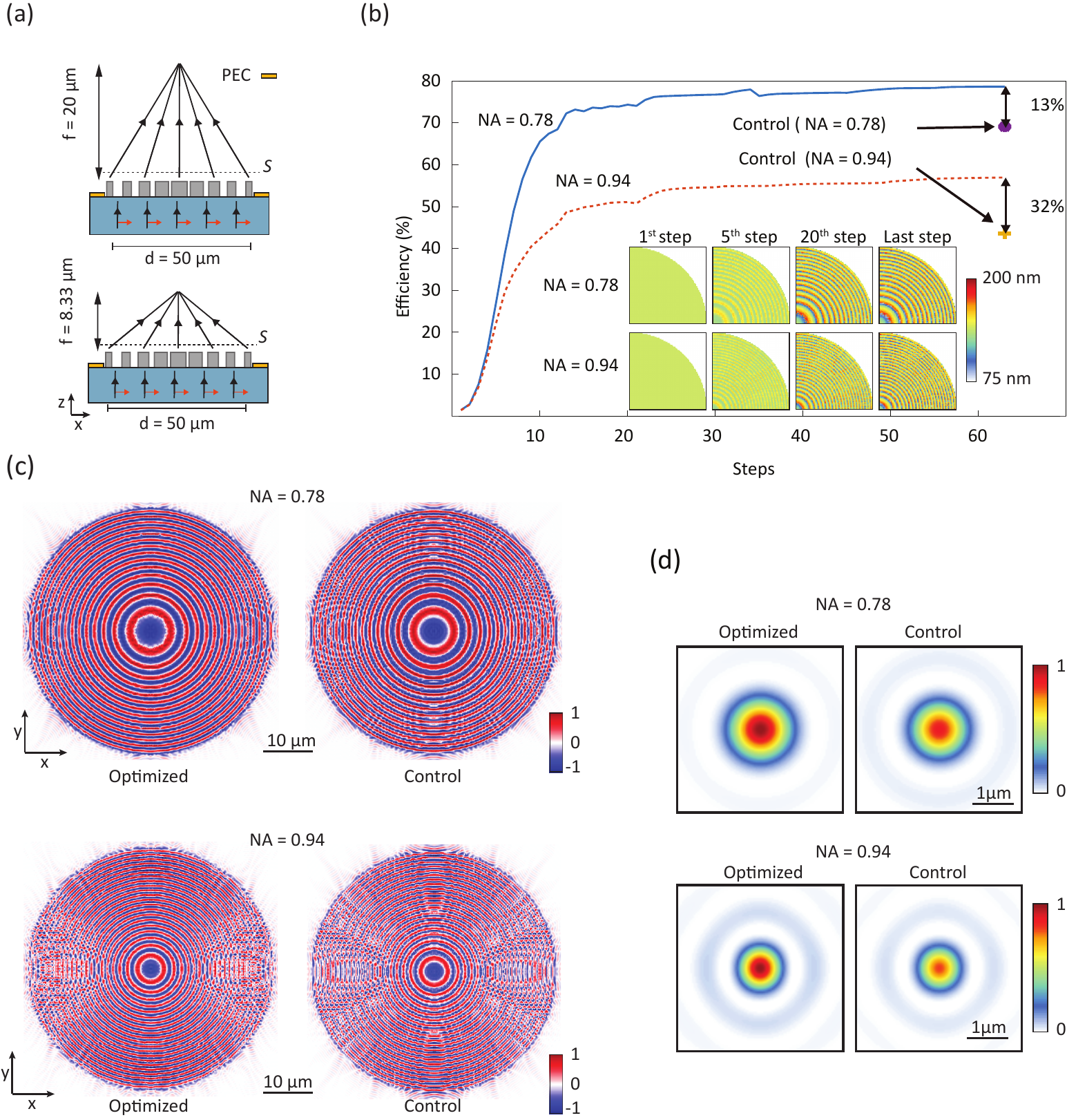}}
	\caption{Simulation and design of the optimized and control metalenses. 
		\textbf{(a)} Schematic of two metalenses with NAs of 0.78 and 0.95. The metalenses are illuminated by normally incident $x$-polarized plane waves. The incident field outside the metalens aperture is blocked by a perfect electric conductor (PEC) layer. 
		\textbf{(b)} The focusing efficiencies of the optimized metalenses during the optimization process. Focusing efficiencies of the control metalenses are shown for comparison. Inset shows color-coded width distributions of meta-atoms at several steps during the optimization process. 
		\textbf{(c)} Snapshot of $E_x$ on the output apertures of the optimized (left) and control metalenses (right). 
		\textbf{(d)} Intensity at the focal planes of the optimized and control metalens.}
	\label{simulations}
\end{figure*}

To demonstrate the parameterized adjoint method, we designed two metalenses with NAs of 0.78 and 0.94 (Fig.~\ref{simulations}a). The diameters of both metalenses are 50 $\upmu$m, yielding focal lengths of 20 $\upmu$m and 8.3 $\upmu$m. 

The metalenses are composed of 430-nm-tall square $\alpha$Si meta-atoms that are arranged on a rectangular lattice with a period of 320 nm. The meta-atoms rest on a fused silica substrate and are surrounded by air. 
For these designs, the parameter vector consists of the meta-atom widths, $\mathbf{p}=(w_1,\ w_2,\ \ldots,\ w_N)$, where N$\approx$19,200 is the number of meta-atoms. By imposing symmetries present in the problem we can reduce the design to 4800 independent variables. Still, the large number of independent variables and the long time required for each simulation precludes a detailed study of the design space.
Both metalenses are initialized by a uniform array of 140-nm-wide meta-atoms (i.e., $w_i$=140 nm).

Both forward and adjoint simulations were performed using a finite difference time domain (FDTD) solver \cite{Oskooi2010} with a sinusoidal excitation that was gradually ramped up.  In the forward simulations, the metalenses were illuminated by an $x$-polarized, normally incident plane wave (Fig.~\ref{simulations}a) with a free-space wavelength of $\lambda_0$=850 nm. The desired output field $\mathbf{E}^\text{d}$ was selected to be the field of an ideal, spherical-aberration-free flat lens (see Methods) \cite{andrew_cleo}$\!$. To expedite the simulations, symmetric boundary conditions were used along both $x$ and $y$ axes, reducing the simulation volume by a factor of four. The simulations were run until the results converged, and then the fields were converted from time to frequency domains using the method of ref. \cite{furse2000faster}$\!$. The fields on the meta-atom side boundaries, necessary to determine $\nabla_\mathbf{p}F$, were interpolated from points on the Yee grid using a bilinear approach (see Methods and Fig.~S4). Further simulation details are described in the Methods section.

In each step of the optimization, the design vector was updated according to $\mathbf{p}^{(n+1)}=\mathbf{p}^{(n)}+s\nabla_{\mathbf{p}^{(n)}}I$, where $s$ is the step size. The step size was chosen to achieve an average meta-atom width change of a few nanometers. As the optimization proceeded, the step size was manually updated, allowing $\mathbf{p}$ to converge (see Methods and Fig.~S5). To enforce polarization insensitivity, we symmetrized the derivatives along $x=y$ plane (see Methods and Fig.~S6).
  
As a quantitative measure of performance, we calculated the focusing efficiency of each metalens during the optimization process. Focusing efficiency is directly related to the accurate implementation of the desired field profile, and metalenses with higher focusing efficiencies generally have less undesired scattered light and form higher contrast images close to their optical axes.
For a fair comparison with the measured values (see the Experimental demonstration section below), we defined the focusing efficiency as the fraction of the power incident on the metalens aperture that passes through a 7-$\upmu$m-diameter aperture in its focal plane.
Figure 4b shows the focusing efficiencies of the optimized metalenses as their design evolved during the optimization process. Color-coded meta-atom width maps for these metalenses at several steps during the design process are shown as insets in Fig.~\ref{simulations}b. At the first step of the optimization, the metalenses were periodic arrays of posts and had low focusing efficiencies. As the design proceeded, patterns similar to Fresnel zones appeared in the metalenses’ width distributions (Fig.~\ref{simulations}b, insets), and their focusing efficiencies increased.

The designs were run for 64 iterations, although after only 25 steps their focusing efficiencies reached plateaus. 
At the last step, the focusing efficiencies of the optimized metalenses with NAs of 0.78 and 0.94 were 78\% and 55\%, respectively. For comparison, we designed two control metalenses using the unit-cell approach with NAs, meta-atom heights, lattice constants, and diameters identical to the optimized ones. The simulated focusing efficiencies of the control metalenses are 69\% and 43\%. The details of the designs and simulations of these control metalenses are presented in Methods. Snapshots of the dominant component of the electric field ($E_x$) at the output apertures of the control and optimized metalenses are presented in Fig.~\ref{simulations}c (for $E_y$ distributions see Fig.~S7). The field distributions in Fig.~\ref{simulations}c show that the optimized metalenses generate the desired fields with smaller phase errors, and consequently produce brighter focal spots than the control metalenses (Fig.~\ref{simulations}d). The significantly higher focusing efficiencies of the optimized metalenses compared to their control counterparts demonstrate the efficacy of the parameterized adjoint optimization technique in designing high-performance metasurfaces.

\subsection*{Experimental demonstration}
\begin{figure*}
	\centerline{\includegraphics{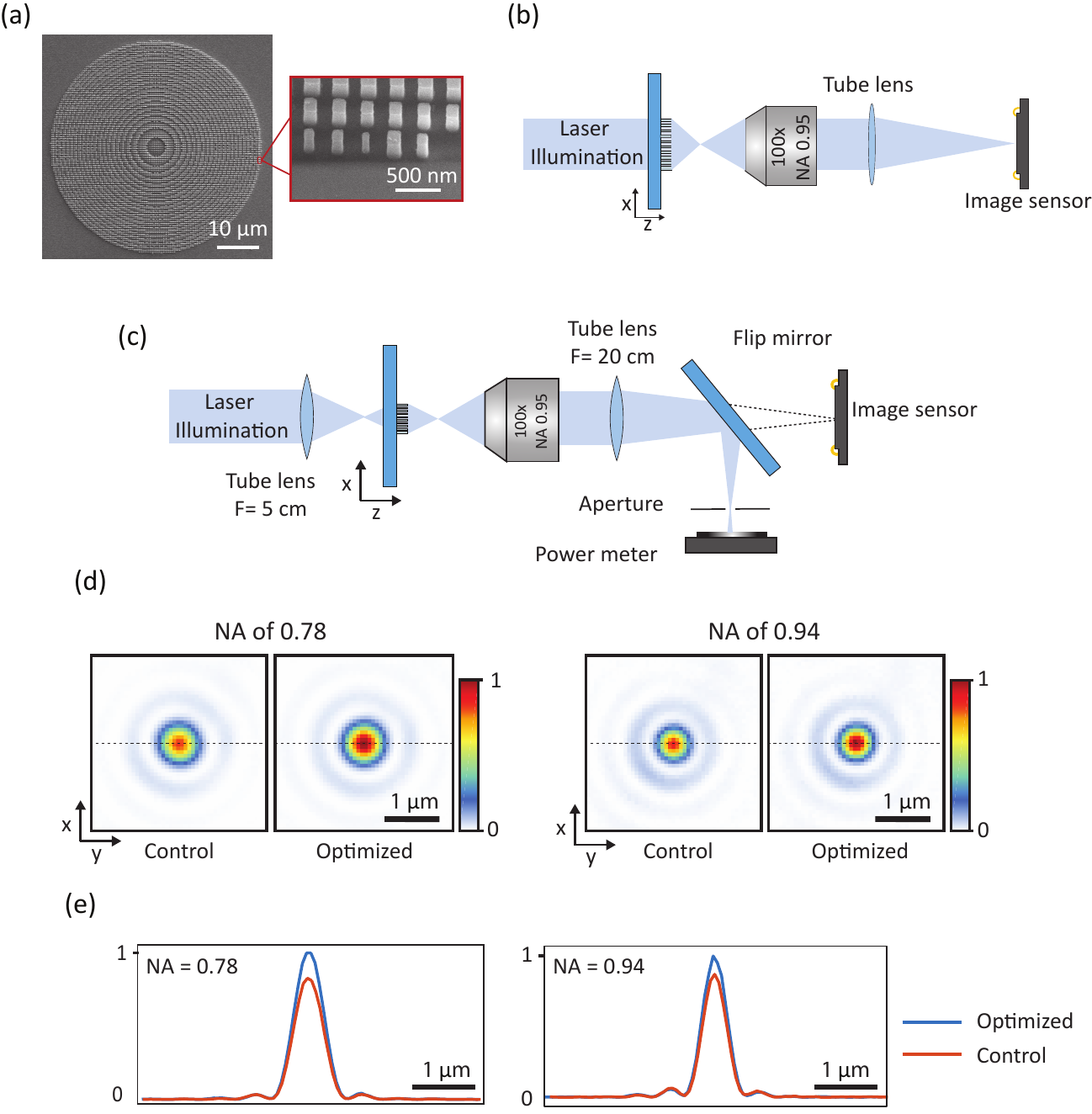}}
	\caption{Experimental results. 
		\textbf{(a)} Scanning electron beam micrograph of a fabricated metalens. 
		\textbf{(b)} Schematic of the characterization setup for intensity measurements and \textbf{(c)} efficiency measurements. 
		\textbf{(d)} Intensity distributions of optimized and conventional metalenses. \textbf{(e)} Intensity profiles are taken along the dashed lines shown in \textbf{(d)}.}
	\label{experimental}
\end{figure*}

For experimental validation, we fabricated and characterized the optimized and control metalenses. The metalenses were fabricated by depositing a layer of aSi on a fused silica substrate and patterning it using electron beam lithography and dry etching (see Methods for details). Figure 5a shows an SEM image of a fabricated metalens. We characterized the metalenses using a setup schematically shown in Fig.~\ref{experimental}b. Metalenses were illuminated by a collimated laser beam with a wavelength of $\lambda_0$=850 nm. The light transmitted through the metalens was collected by an objective lens with an NA of 0.95 and reimaged by a tube lens on an image sensor. Images of the focal spots, shown in Fig.~5b, show enhanced peak intensities for the optimized metalenses compared to the control ones.

We measured the focusing efficiencies of the metalenses by measuring the ratio of the optical power focused into a 7-$\upmu$m-diameter pinhole in the focal plane of the metalenses and the power incident on their apertures (Fig.~5c). The measured focusing efficiencies of the optimized metalenses with NAs of 0.78 and 0.94 are 65\% and 49\%, respectively, higher than values of 52\% and 43\% obtained for their control counterparts. This represents 24\% and 13\%  relative enhancements for the 0.78 and 0.94 NA lenses, respectively. The smaller increase for the higher NA metalens is attributable to the limitations of our measurement setup (the objective lens used has an NA of 0.95) and to its higher sensitivities to fabrication errors.

To study the sensitivity of our designs, an array of metalenses with a range of constant meta-atom offsets were fabricated alongside those characterized in Fig.~ \ref{experimental}. The study shows that the optimized metalenses have approximately the same sensitivities as the control ones (see Fig.~S8).

\section*{Discussion}

The parameterized adjoint optimization method accurately estimates shape derivatives of parameterized meta-atoms (see Supplementary Note 5 and Fig.~S9). In contrast with methods that simulate structures in a dielectric continuum and then discretize to obtain a physically realizable design \cite{Jensen2011, su2018inverse, sell2017, sell2017periodic}, meta-atoms designed by our method maintain a dielectric discontinuity at their boundaries throughout the whole design process, i.e., the simulation and design domains are the same. 
Techniques such as level-set representation can also be used to maintain boundaries with a dielectric discontinuity. We previously demonstrated such a technique in a similar silicon on glass material platform \cite{Mahdad_levelset}. Compared to the parametrized technique presented in this article, the simulations for the free-form level-set technique require significantly higher resolutions (i.e., much smaller grid size) to converge and the optimization domain has many more local optima. Due to their small features, the optimized metasurfaces obtained using this level-set approach are also significantly more difficult to fabricate. As a result, the application of level-set representation has been limited to small structures\cite{Mahdad_levelset}.

The parameterized adjoint optimization technique can be easily adapted for designing other types of metasurfaces such as achromatic metasurfaces (see Supplementary Note 4). 
We have presented the design of achromatic metalenses with parameterized shapes in Figs.~S10 and S11.
These metasurfaces provide comparable efficiencies to the ones designed using topology optimization \cite{Chung2020highNA}, and do not pose fabrication challenges similar to those of free-form structures.

Using simple, parameterized shapes reduces the dimensionality of the metasurface design space and simplifies the fabrication process.
Designs produced by adjoint topology optimization typically require hundreds of steps to converge \cite{sell2017, Jensen2011}.  
Parametrization enables us to include our knowledge about principles of operation of metasurfaces by selecting proper arrangement of the meta-atoms and other parameters such as meta-atom height and lattice constant. Our initial design ( uniform metasurface comprising identical meta-atoms) although very simple, includes many important characteristics of the final design, so it can converge faster. The metalenses presented in this work evolved to designs with performance superior to the conventionally-designed controls in fewer than 15 steps.
The quick convergence enabled us to optimize large-scale (50 $\upmu$m diameter) metastructures, which, to the best of our knowledge, are currently some of the largest 3D adjoint-optimized metalenses. 
We previously demonstrated multifunctional multi-layer metasurface ref. \cite{Mahdad_optica} devices using similar methods in approximately the same number of iterations. 
Furthermore, the number of iterations could be further reduced by implementing an adaptive step size \cite{johnson2014nlopt}.

The full-wave simulations employed in this work are computationally expensive. We employed several techniques to keep the optimization of large devices feasible. The computational cost of FDTD simulation is directly related to the grid size used. We employed bi-linear field interpolation, which increases the accuracy of the derivatives without reducing the grid size, keeping the computation time for each iteration tenable. To convert the time-domain fields to the frequency domain, we only used two time samples using an efficient harmonic detection method \cite{furse2000faster}. This technique enables multi-wavelength optimization at minimal additional cost (see Supplementary Note 4): wavelengths with independent objective functions can be incorporated into the simulations by adding appropriate sources and acquiring a few additional time-domain samples without increasing the number or duration of the simulations.
Though in this work we presented metalenses optimized from a trivial initial state, we could have selected a conventionally designed metasurface (based on a unit-cell approach) as a starting point, which might have positioned the initial and final designs nearer to each other. 

Like any other gradient-based optimization method, designs determined by our method represent local optima. 
However, parameterization allows us to restrict our search to a judiciously selected subspace by using prior knowledge about the problem. For example, information from low NA conventional designs can be useful in determining the appropriate meta-atom height and the lattice constant for a high NA adjoint-optimized design.
To improve the chance of finding the global optimum, multiple optimizations starting from initial designs can be run in parallel. Results of such a multiple-seed optimization are shown in Fig.~S12. Despite their different starting points, all designs converged to metalenses with similar focusing efficiencies. The observed behavior might not be general, but it seems to be valid at least for optimizing single layer structures with significant practical impact.

Because our method requires little knowledge about the final structure, it allows us to design elements for which conventional techniques fail to produce efficient designs, like multifunctional metasurfaces \cite{Mahdad_optica, Arbabi2016a_multiwavelength, kamali2017angle, zhou2018multilayer}. In multifunctional devices, the interdependence of parameters is significantly more complex than in single function designs and simple models are unable to model meta-atom behavior accurately.  In contrast, our method considers all the complex interactions and generates more efficient designs. Our method is also can be easily extended to other kinds of multi-objective optimizations, like robust designs, that are tolerant to fabrication \cite{oskooi2012robust} error. 

We envision that the adaptation of the parameterized adjoint optimization to design of large-scale metasurfaces will enable efficient cascading of multiple metasurfaces to implement compact, complex metasystems with high performance.

\begin{acknowledgement}
This work was funded by the Samsung Advanced Institute of Technology, and performed in part at the Center for Nanoscale Systems (CNS) at Harvard University, a member of the National Nanotechnology Coordinated Infrastructure Network (NNCI), which is supported by the National Science Foundation under NSF award no. 1541959.
\end{acknowledgement}

\section*{Methods}

\subsubsection*{Metalens optimization}
The two metalenses designed by the adjoint technique and the control metalenses are composed of 430-nm-tall square cross-section aSi meta-atoms ($n_\text{Si}=3.84$) that are positioned on a square lattice with a lattice constant of $\Lambda=320$~nm. 
The meta-atoms are on a fused silica substrate ($n_\text{s}=1.45$) and are cladded above by vacuum.
One quadrant of each of the metalenses are shown in Fig.~S7.

The optimization flowchart is shown in Fig.~S3.
To reduce the required computational resources, we simulated the fields in a small volume (52~$\upmu$m $\times$ 52 $\upmu$m $\times$ 1.33~$\upmu$m) around the metasurface.
All metalens optimization simulations were performed using a freely-available, open-source FDTD solver~\cite{Oskooi2010}.
Time-domain simulations were run until the fields converged (133~fs).
The structure is terminated on all sides by a PML boundary condition.
Because only the near field of the structure was simulated, fields at the focal plane (Fig.~4) were obtained by Fourier domain propagation.
To further expedite the simulations, we exploited symmetries of the structure and fields: even mirror symmetry was specified along the $x$-axis and odd mirror symmetry along the $y$-axis, reducing the simulated volume by a factor of four.

Simulations were done using a workstation with an Intel E5-2680 CPU; 10 cores were used for each simulation.
The FDTD grid size and the step size were adjusted manually when a reduction in the rate of improvement was observed.
The simulations in each optimization run began with a grid size of 33~nm (low resolution); after the device efficiency increased, the grid size was reduced to 20~nm (high resolution).
Each iteration, consisting of both forward and adjoint simulations, took $\sim$15~min at low resolution and $\sim$97~min at high resolution.
Color-coded plots of meta-atom widths of the optimized and control lens are shown in Fig. S13.

\subsection*{Target field distribution}
For an $x$-polarized plane wave input $\mathbf{E}^\text{i}=\hat{x}E_0$ at wavelength $\lambda_0$ originating in a medium with refractive index $n_\text{c}$, the desired field distribution for an ideal metalens with focal length $f$ is:
\begin{align}
E_x^\mathrm{d} &= E_0\,t(\theta)\left(
\cos\theta\cos^2\phi + \sin^2\phi
\right)\\
E_y^\mathrm{d} &= E_0\,t(\theta)\left(
1 + \cos\theta
\right)\sin\phi\cos\phi,
\end{align}
where $t(\theta)=\sqrt{\frac{n_\text{c}}{\cos\theta}}\exp(-\frac{2\pi j f}{\lambda_0\cos\theta})$, $\theta=\tan^{-1}\left(\sqrt{x^2 + y^2}/f\right)$\cite{andrew_cleo} is the local deflection angle of the metasurface, and $\phi=\tan^{-1}\left(y/x\right)$ (see Fig.~S14).

\subsection*{Field interpolation}
The FDTD solver calculates fields on a rectangular grid (Yee grid).
However, to determine the gradient, fields on the meta-atom boundaries are required.
From the boundary conditions, we know the fields $D_\perp$ and $\mathbf{E}_\parallel$ are continuous.
To obtain the boundary fields, we interpolated along axes normal to meta-atom boundaries using a two-sided linear fit approach that considers field values at four Yee lattice points (Fig.~S4).
For each field component $C$, one linear fit $C_\text{in}(x) $ was determined using two points $(x_{-2}, x_{-1})$ inside the meta-atom, and another, $C_\text{out}(x) $, using two points $(x_{1}, x_{2})$ outside the meta-atom.
The field at the boundary ($x_0$) was found based on the distance-weighted average of these two extrapolated values as

\begin{equation}
C(x_0) \approx \alpha C_\text{out}(x_0) + \beta C_\text{in}(x_0),
\end{equation}
where $\alpha$ and $\beta$ are given by
$$ \alpha = \frac{\left| x_{0} - x_{-1} \right|}{\left|x_{1} - x_{-1}\right|},\beta = \frac{\left| x_{0} - x_{1} \right|}{\left|x_{1} - x_{-1}\right|}.$$

\subsection*{Gradient symmetrization and scaling}
To obtain polarization-insensitive metalens designs, in addition to the mirror symmetries along $x$ and $y$ axes described above for the simulation domain, we imposed a symmetry along the $x=y$ line (see Fig.~S6).
The gradients were first determined for the simulated, $x$-polarized field for a quarter of the metalens and then symmetrized according to:
\begin{equation}
\nabla_\mathbf{p}I(x, y) \leftarrow \frac{1}{2}\nabla_\mathbf{p}I(x, y) + \frac{1}{2}\nabla_\mathbf{p}I(y, x).
\end{equation}
This operation is equivalent to computing the gradient for circularly polarized input light and optimizing the metalens using this symmetrized gradient ensures its polarization insensitivity.
After determining the symmetrized gradient, the step size $s$ was selected such that the average of the absolute change of the meta-atom widths $\nabla_\mathrm{p} I$ was equal to a few nanometers (see Fig.~S5). The maximum change in the post widths was limited to 10 times the average value to ensure the first order gradient approximation is valid.

At the beginning of the optimization the absolute value of the average change was selected to be equal to 2~nm.
Then, as the reduction in the rate of improvement was observed (see Fig.~S5), it was reduced to 0.1 $\text{nm}$.

\subsection*{Control metalens designs}
To compare the effectiveness of the proposed design method with the conventional unit-cell design approach, we designed two control metalenses using the unit-cell approach.
The control metalenses have the same design parameters as the optimized ones, i.e., with lattice constants of 320~nm, and square cross-section aSi meta-atoms ($n_\text{Si}=3.84$) that are 430~nm tall.
Simulated transmittance and phase of the transmission coefficient  for a periodic array of meta-atoms are shown in Fig.~S15a and were used to obtain the design map shown in Fig.~S15b.

\subsection*{Fabrication}
All metalenses were fabricated on the same fused silica substrate.
To compensate for systematic errors in lithography, etching and other fabrication processes, an array of offsetted designs were included in the pattern. 
In each element of this array, widths of square meta-atoms are uniformly changed by a value in a range of $-$15 nm to 45 nm in steps of 5 nm. 
Figure S8 shows the measured efficiencies of fabricated metalenses with different offset values.

To pattern the metasurfaces, a 430-nm-thick layer of aSi was deposited on the substrate using plasma-enhanced chemical vapor deposition.
Then, an approximately 220-nm-thick layer of electron-beam resist (ZEP520A-7, Zeon) was spin coated on the substrate.
To avoid charging effects, a conductive polymer layer (ARPC-5090, Allresist) was spin coated on top of the resist.
The patterns were defined using a 125~kV electron-beam lithography system (ELS-F125, Elionix), and then an aluminum oxide hard mask was deposited using an electron-beam evaporator.
After lifting off the hard mask in a solvent (Remover PG, Microchem), the sample was etched using an inductively-coupled plasma reactive ion etching tool in an SF$_{6}$/C$_{4}$F$_{8}$ gas mixture.
The hard mask was removed in a heated solution of ammonium hydroxide and hydrogen peroxide.

\subsection*{Characterization}
We used the setup schematically drawn in Fig.~S16a to acquire the focusing efficiency of the metalenses.
Each metalens was illuminated by a weakly diverging Gaussian beam with a wavelength of 850~nm that was partially focused by a lens with 5~cm focal length (AC254-050, Thorlabs).
The light passed through the metalens and came into focus at a focal plane.
Light in the focal plane was collected by a microscope objective with an NA of 0.95 (UMPlanFI 100$\times$, Olympus), and reimaged by a tube lens (AC254-200, Thorlabs) and a camera (CoolSnap K4, Photometrics).

The focusing efficiency was defined as the ratio of the power focused inside a 7-$\upmu$m-diameter pinhole in the focal plane of the metalens to the total power incident on the metalens.
To measure the efficiency, we measured the power in the reimaged focal plane passing through an aperture with a diameter equivalent to a 7-$\upmu$m-diameter pinhole in the metalens focal plane.
A flip mirror in the imaging system (dashed box in Fig. S16) allowed us to direct the reimaged spot toward an apertured power meter (S120C, Thorlabs) and measure the focusing efficiencies.

The total incident power is measured by redirecting all the power in the partially focused beam (Fig.~S16b) to the power-meter.
Power incident on the metalens is different from the the measured power because of the reflection at the second interface of fused silica to air (Fig.~S16b).
The measured power was corrected to indicate the actual power incident on the lens.

\bibliography{References}

\end{document}


\clearpage
\maketitle
\renewcommand{\theequation}{S\arabic{equation}}
\setcounter{equation}{0}

\section*{Supplementary Note 1: Sensitivity of metasurfaces to subwavelength geometric changes}
\begin{figure*}[h!]
	\centering
	\includegraphics[width=5in]{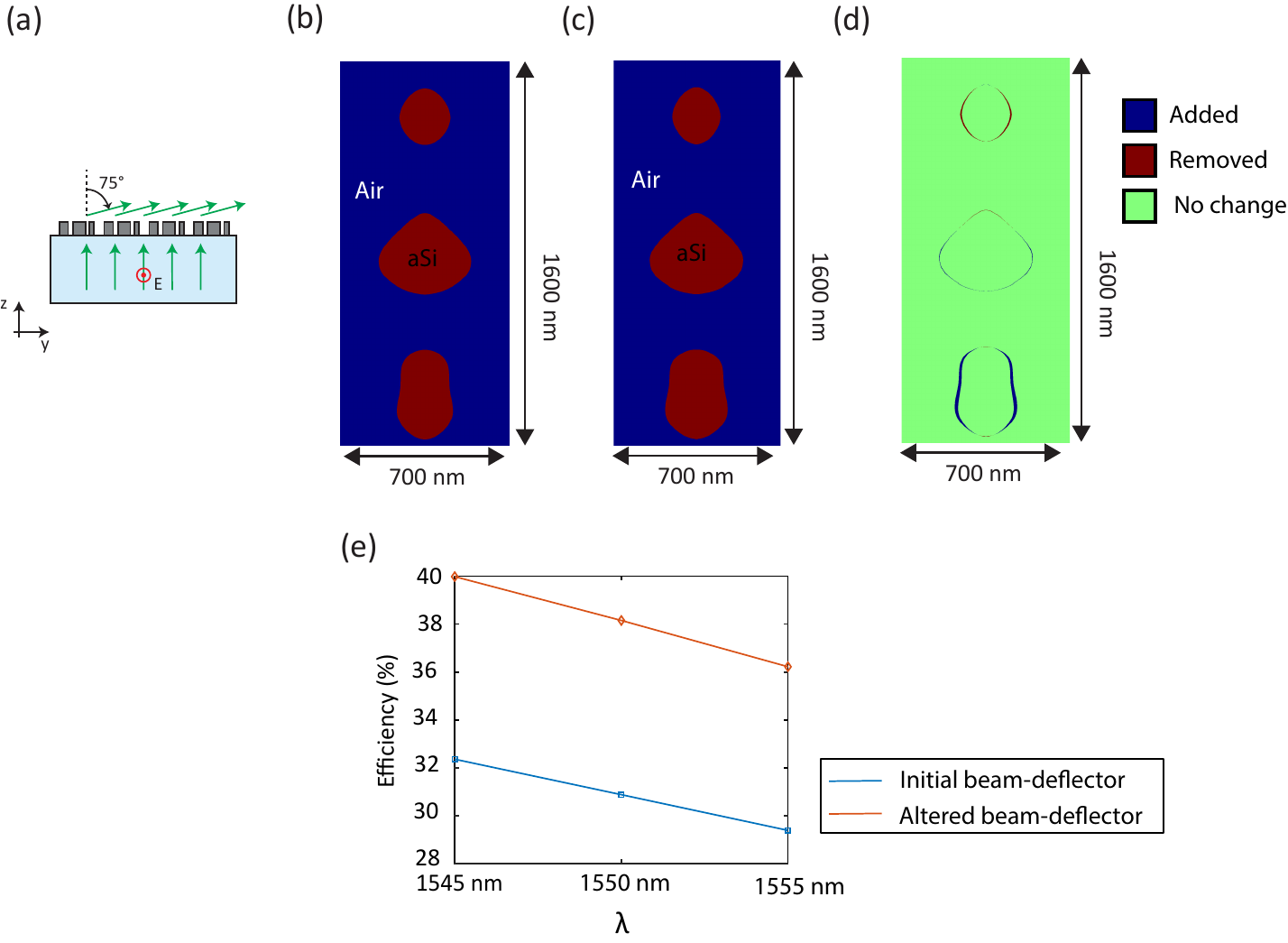}
	\caption{
		Freeform beam deflectors.
		\textbf{(a)} Schematic of the beam deflector.
		\textbf{(b)} Initial beam-deflector.
		\textbf{(c)} Beam-deflector after applying small boundary changes on the meta-atoms. 
		Red areas in \textbf{(b)} and \textbf{(c)} indicate $\alpha$-Si and blue areas indicate air.
		\textbf{(d)} The difference between two designs in \textbf{(b)} and \textbf{(c)}.
		Blue indicates the areas where $\alpha$-Si is added and red where it is removed.
	\textbf(e) Deflection efficiencies before and after applied changes. } 
	\label{freeform}
\end{figure*}

To show that subwavelength features can have a strong effect on metasurface performance, we compare the deflection efficiencies of two very similar beam deflectors.
Fig.~S\ref{freeform}a shows a schematic of the beam deflector: The elements lie on  a rectangular lattice with lattice constants of $a_x=700$~nm and $a_y=1600$~nm.
The heights of the posts are chosen to be $h=800$~nm and they are made of $\alpha$-Si with refractive index of $n=3.6$.
The posts are surrounded by air ($n=1.0$) and rest on a fused silica substrate ($n=1.456$).
The beam deflectors are designed to deflect a normally-incident, $x$-polarized field at a wavelength $\lambda=1550$~nm to an angle of 75$^\circ$.

Figure S\ref{freeform}b and c are both beam deflector designs and very similar to each other: Their boundaries change only by ca.~10~nm (See Fig. S\ref{freeform}d).
However, the efficiency of these devices is quite different: the design in Fig.~S\ref{freeform}b has an efficiency of 30\%, whereas the design in Fig.~S\ref{freeform}c has an efficiency of 38\%.

\clearpage
\maketitle
\renewcommand{\theequation}{S\arabic{equation}}
\setcounter{equation}{0}

\section*{Supplementary Note 2: Comparison of subpixel averaging for free-form and parameterized shapes}
Methods such as subpixel smoothing \cite{farjadpour2006improving, Oskooi2010} can be used to increase the accuracy of simulations without resorting to a high-resolution grid.
In most simulations presented in this manuscript we used MEEP~\cite{Oskooi2010}, which implements subpixel smoothing by default.
Subpixel smoothing can be applied to structures with any shape, however, the computational cost of performing subpixel averaging for free-form structures is considerably higher than applying the same method for shapes with simpler boundaries (like the rectangular meta-atoms used in this work).

\begin{figure*}
	\centering
	\includegraphics[width=5in]{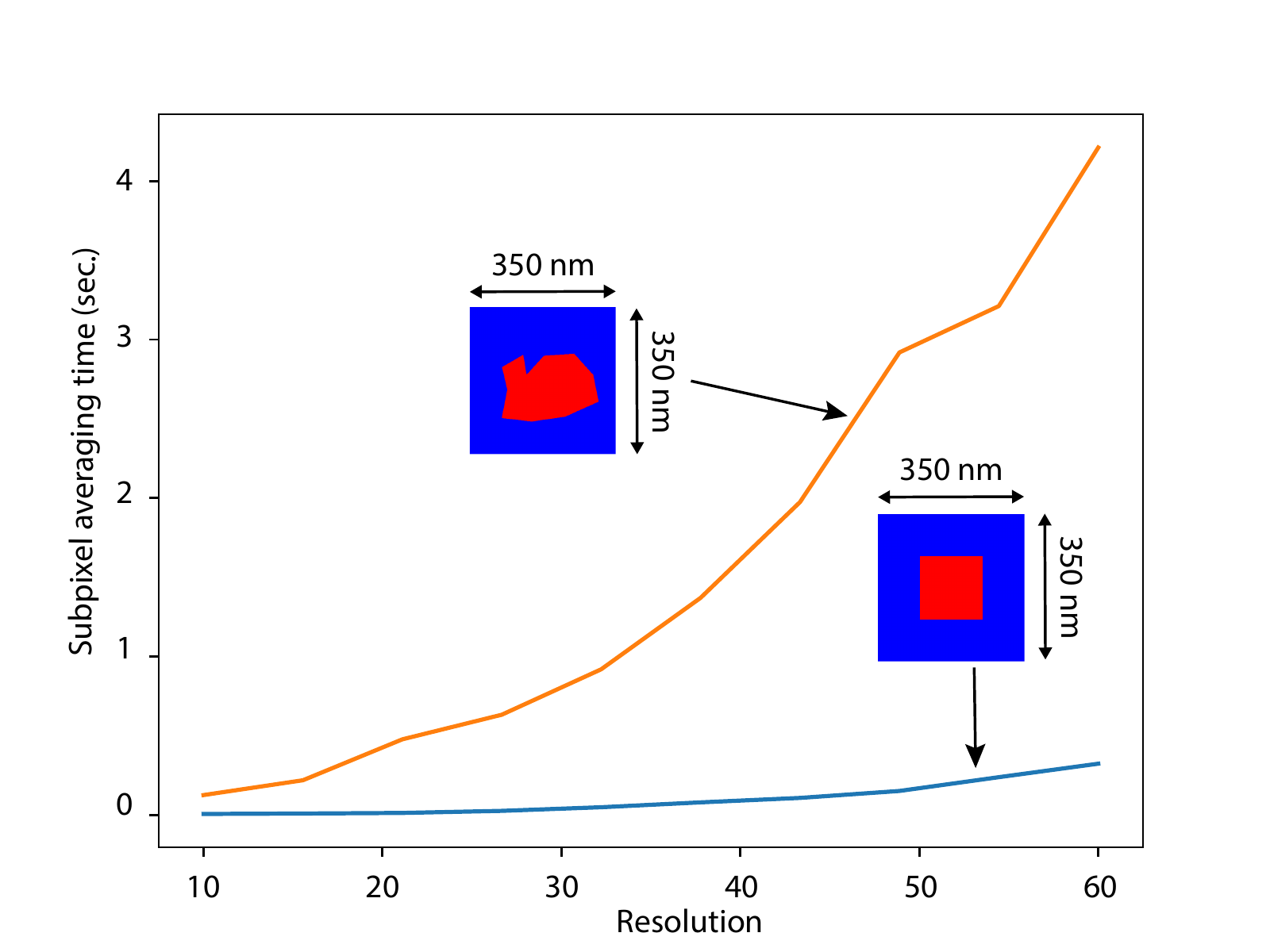}
	\caption{
		Comparison between subpixel smoothing time for a free-form (orange) and parameterized (blue) meta-atom.
	}
	\label{subpixel}
\end{figure*}

We performed a study of the time taken to perform subpixel smoothing as a function of resolution for two representative shapes: the first is a nano-post with a rectangular cross-section, and the second is a nano-post with a more complicated polygonal cross-section. The unit cell for both structures is the same, comprising a fused silica substrate with $n=1.456$ and cladded with air.
The meta-atoms are 500~nm tall and are made of $\alpha$-Si with $n=3.824$ and the simulations wavelength is selected to be 850~nm. The time to initialize each structures using 4 threads as a function of resolution is shown in Fig.~S\ref{subpixel}.
As the Figure S\ref{subpixel} clearly shows, the time to initialize the polygonal shape is considerably higher than that of the rectangular shape.

\clearpage
\maketitle
\renewcommand{\theequation}{S\arabic{equation}}
\setcounter{equation}{0}

\section*{Supplementary Note 3: Gradient of complex objective}
Here we derive Eq.~(2), the expression for the gradient of the complex projection $F$ using the Lorentz reciprocity theorem.
The complex projection $F$ for a given design $\mathbf{p}$ is determined by Eq.~(1).
A small change in $\mathbf{p}$, understood to be the differential increase $\delta w$ in the size of one of the meta-atoms, produces a related design $\mathbf{p}'$ and an associated complex projection $F'$.
The difference between these two designs, $\delta F\equiv F' - F$, is given by
\begin{equation}
  \delta F = \int_S\mathbf{E}^{\mathrm{d}*}\cdot\delta\mathbf{E}^\mathrm{f}\,\mathrm{d}A,
  \label{eq:deltaf}
\end{equation}
where $\delta\mathbf{E}^\mathrm{f}\equiv\mathbf{E}^\mathrm{f}(\mathbf{p}')-\mathbf{E}^\mathrm{f}(\mathbf{p})$ is the difference between the forward fields realized by these two designs.

The Lorentz reciprocity theorem specifies a relationship between two sources at the same frequency $\omega$, $\mathbf{J}_1$ and $\mathbf{J}_2$, and the fields produced by them, $\mathbf{E}_1$ and $\mathbf{E}_2$, in the same linear, reciprocal environment~\cite{Harrington2001}:
\begin{equation}
  \int\mathbf{E}_1\cdot\mathbf{J}_2\,\mathrm{d}V = \int\mathbf{E}_2\cdot\mathbf{J}_1\,\mathrm{d}V.
  \label{eq:reciprocity}
\end{equation}
Here the integral is understood to extend over all space.
Though $\mathbf{p}$ and $\mathbf{p}'$ represent \emph{different} media, $\mathbf{E}^\mathrm{f}(\mathbf{p}')$ can be understood as the sum of $\mathbf{E}^\mathrm{f}(\mathbf{p})$ and fields by an equivalent polarization current

\begin{equation}
  \mathbf{J}_\text{eq}(\mathbf{r}) = j\omega\left(n^2_\mathrm{m}-n^2_\mathrm{c}\right)\mathbf{E}^\mathrm{f}(\mathbf{r}, \mathbf{p}'),
  \label{eq:current_equivalent}
\end{equation}
where $\mathbf{r}$ is in the region formed of $n_\text{m}$ in $\mathbf{p}'$ that previously consisted of $n_\text{c}$ in $\mathbf{p}$, and $n_\text{m}$ and $n_\text{c}$ are the refractive indices of the meta-atom and the cladding material, respectively.
We will label this region $\chi$.
Thus current $\mathbf{J}_\text{eq}$ and the fields $\delta\mathbf{E}^\mathrm{f}$ form one of the source--field pairs in Eq.~\ref{eq:reciprocity}.
To determine the value of $\delta F$, we can consider a current source $\mathbf{J}^\mathrm{a}_\mathrm{s}\equiv \mathbf{E}^{\mathrm{d}*}$ on $S$ and its associated response $\mathbf{E}^\mathrm{a}(\mathbf{p})$.
Using Eqs.~\ref{eq:reciprocity} and \ref{eq:current_equivalent}, we can rewrite Eq.~\ref{eq:deltaf} as
\begin{equation}
  \delta F
  = j\omega(n^2_\mathrm{m}-n^2_\mathrm{c})\int_\chi\mathbf{E}^\mathrm{a}(\mathbf{p})\cdot\mathbf{E}^\mathrm{f}(\mathbf{p}')\,\mathrm{d}V.
\end{equation}

We would like to determine $\delta F$ without simulating the perturbed geometry $\mathbf{p}'$, so we must approximate $\mathbf{E}^\mathrm{f}(\mathbf{p}')$ using known quantities.
We will do this using the dielectric boundary conditions:
The components of the electric field parallel to a dielectric boundary, $\mathbf{E}_\parallel$, are continuous.
Similarly, the component of the displacement field perpendicular to a dielectric boundary, $D_\perp=\epsilon E_\perp$, are continuous.
For a small change in the meta-atom geometry, we can make the approximations
\begin{align}
  \mathbf{E}_\parallel(\mathbf{p}') &\approx \mathbf{E}_\parallel(\mathbf{p}),\\
         {D}_\perp(\mathbf{p}') &\approx {D}_\perp(\mathbf{p}).
\end{align}

The integration domain $\chi$ between $\mathbf{p}$ and $\mathbf{p}'$ follows the contour of the meta-atom $\partial\Omega$ and has thickness $\delta w$; thus

\begin{align}
  \delta F &= j\omega(n^2_\mathrm{m}-n^2_\mathrm{c})
  \int_{\partial\Omega}
  \left[\mathbf{E}_\parallel^\mathrm{a}(\mathbf{p})\cdot\mathbf{E}_\parallel^\mathrm{f}(\mathbf{p})
    + \frac{1}{n^2_\mathrm{m}n^2_\mathrm{c}}{D}_\perp^\mathrm{a}(\mathbf{p}) {D}_\perp^\mathrm{f}(\mathbf{p})
    \right]\,\mathrm{d}A\,\delta w/2,
\end{align}
giving the expression in Eq.~(2).


\clearpage
\maketitle
\renewcommand{\theequation}{S\arabic{equation}}
\setcounter{equation}{0}

\section*{Supplementary Note 4: Multi-wavelength optimization}
We present a straightforward extension of our parameterized adjoint optimization technique to perform multi-wavelength optimization.
Instead of a single harmonic excitation sources, for multi-wavelength optimization, a multi-tone source which includes all the desired wavelengths is used.
Thus adding additional wavelength objectives requires almost no additional computational resources.

To calculate the objective function and the boundary fields on the meta-atoms we sample the fields at more time step and  calculate the contributions of each distinct wavelengths.
At each wavelength, the gradient is calculated similarly to single wavelength case explained in Supplementary Note 3.
For multi-wavelength optimization, the total objective function $I_\text{tot}$ is chosen to be the product of each wavelength component:
\begin{equation}
	I_\text{tot} = I_{\lambda_0} \ I_{\lambda_1} I_{\lambda_2} \cdots
\end{equation}
The gradient of $I_\text{tot}$ is obtained via the product rule:
\begin{equation}
	\nabla_\mathbf{p}I_\text{tot} = (\nabla_\mathbf{p}I_{\lambda_0}) I_{\lambda_1} I_{\lambda_2} \cdots +
	I_{\lambda_0} (\nabla_\mathbf{p}I_{\lambda_1}) I_{\lambda_2} \cdots + \cdots
\end{equation}

\begin{figure*}
	\centering
	\includegraphics[width=5in]{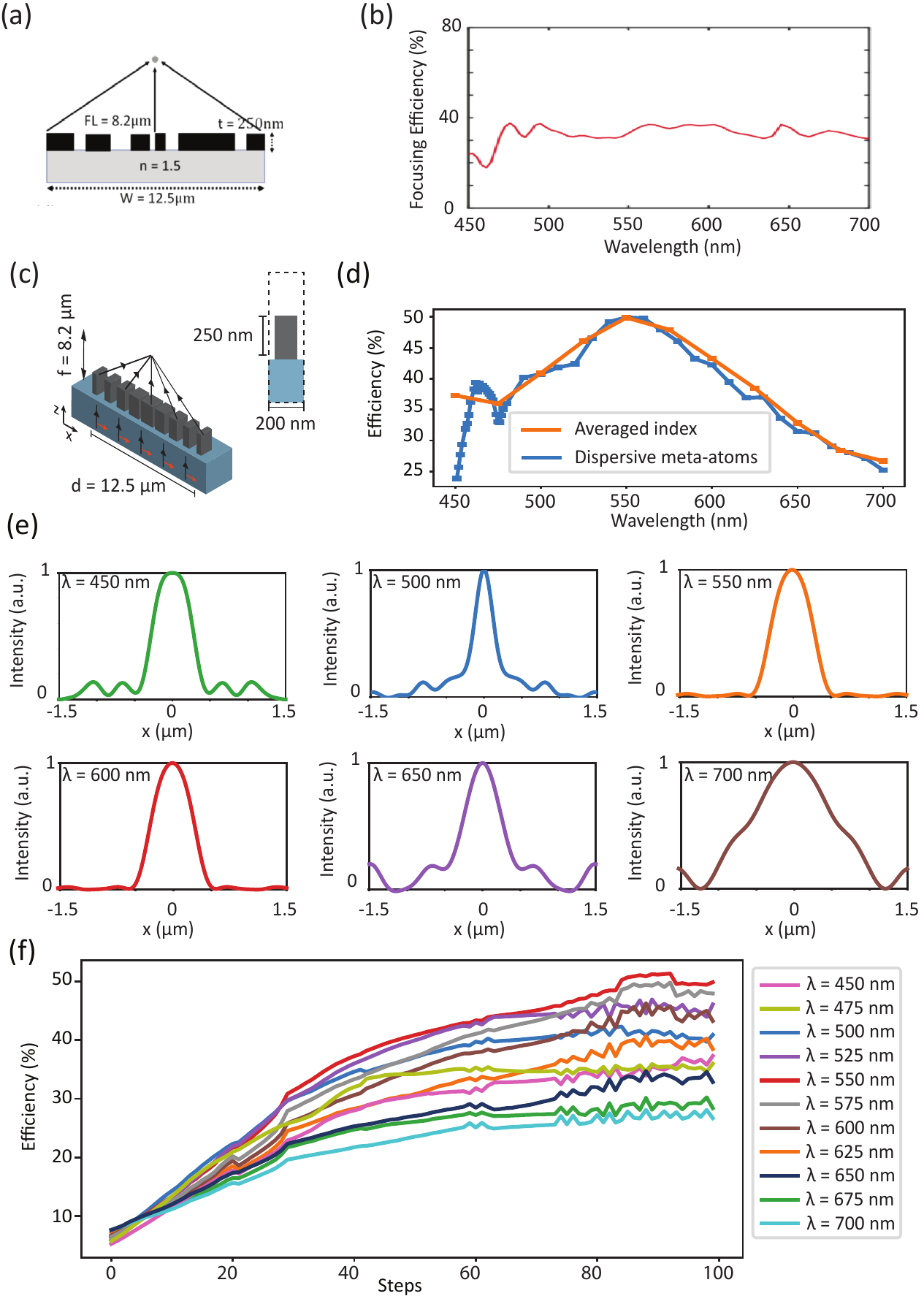}
	\caption{
		Comparison of 250-nm-thick multi-wavelength cylindrical lenses with NA=0.6 designed with Ref. \cite{Chung2020highNA} and the parametrized adjoint optimization techniques.
		\textbf{(a)} Schematic drawing and \textbf{(b)} focusing efficiencies of topology optimized two-dimensional lens adapted with permission from \cite{Chung2020highNA}.
		\textbf{(c)} Schematic drawing of the cylindrical lens and its unit-cell designed with adjoint optimization method.
		\textbf{(d)} Focusing efficiencies.
		\textbf{(e)} Normalized field-intensity profiles at each wavelength.
		\textbf{(f)} Efficiencies at each step of the optimization.}
	\label{Multiwavelength}
\end{figure*}

We present two multiwavelength cylindrical lenses with materials, layer heights and objectives similar to designs presented in Ref.~\cite{Chung2020highNA} (Fig.~S\ref{Multiwavelength} a-b), allowing our technique to be compared with topology-optimized designs.
We optimized a 12.5-$\upmu$m-wide cylindrical metasurface shown in Fig.~S\ref{Multiwavelength}c.
The metasurface is composed of an array of titanium dioxide meta-atoms ($n=2.645$) with height $h=$250~nm (see inset of Fig.~S\ref{Multiwavelength}a) arranged on a rectangular lattice with a lattice constant of 200~nm.

The design is optimized at 11 wavelengths from 450~nm to 700~nm in steps of 25~nm.
The metasurface is excited by an $x$-polarized incident field and designed to focus light at a focal distance $f=$8.2~$\upmu$m (i.e., NA=0.6).
The optimization proceeds for 99 steps starting from an initial design in which all meta-atoms have equal width $w=$140~nm.
The normalized intensity at the last step is shown in Fig.~S\ref{Multiwavelength}e.

Initially all the meta-atoms have square cross-section.
Over the course of the optimization the cross-sections are allowed to change to be rectangular; meta-atoms are also free to move in the $x$ direction.
The efficiency of the metasurface at each optimized wavelength over the course of the optimization is shown in Fig.~S\ref{Multiwavelength}d.
Here the efficiency is defined as the ratio of the power between the first zero of an ideal diffraction limited lens on either side of maximum intensity to the power incident on the lens from substrate.

\begin{figure*}
	\centering
	\includegraphics[width=5in]{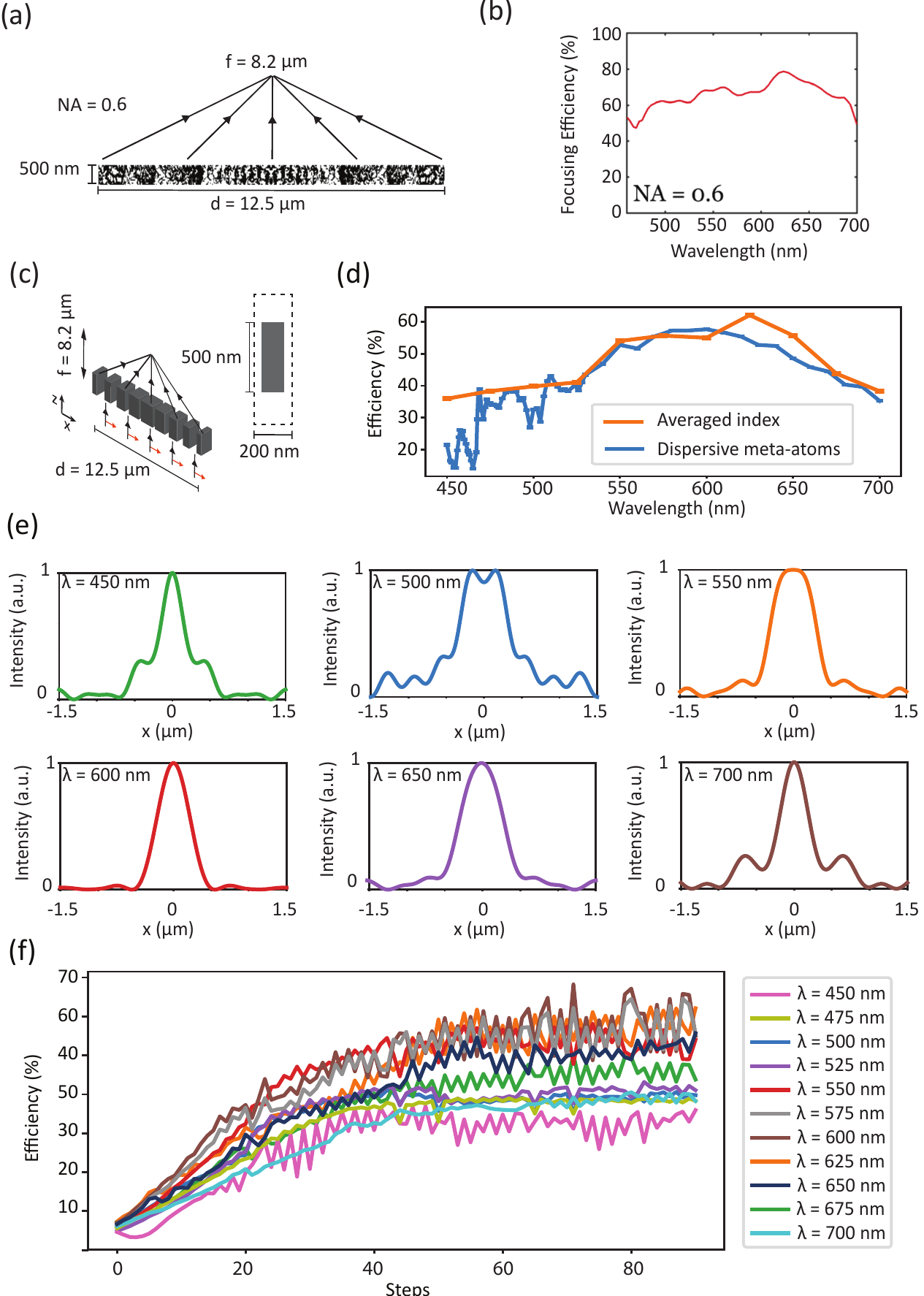}
	\caption{
		Comparison of 500-nm-thick multi-wavelength cylindrical lenses with NA=0.6 designed with Ref. \cite{Chung2020highNA} and the parametrized adjoint optimization techniques.
		\textbf{(a)} Schematic drawing and \textbf{(b)} focusing efficiencies of topology optimized two-dimensional lens adapted with permission from \cite{Chung2020highNA}.
		\textbf{(c)} Schematic drawing of the cylindrical lens and its unit-cell designed with adjoint optimization method.
		\textbf{(d)} Focusing efficiencies.
		\textbf{(e)} Normalized field-intensity profiles at each wavelength.
		\textbf{(f)} Efficiencies at each step of the optimization.}
	\label{Multiwavelength_tall}
\end{figure*}

A second design is presented in Fig.~S\ref{Multiwavelength_tall}.
In this design we increased the height of the meta-atoms to 500~nm to correspond to the layer height of the freeform design in Ref.~\cite{Chung2020highNA} (See Fig.~S\ref{Multiwavelength_tall} a-b).
The metasurface and its unit-cell is schematically shown in Fig.~S\ref{Multiwavelength_tall}c.
The final efficiencies and the progress of the optimization is shown in Fig.~S\ref{Multiwavelength_tall}d and f.

Material dispersion was not considered in the optimization process and the device was optimized at 11 different wavelengths with a constant refractive index of 2.654 for TiO$_{2}$. To evaluate the effect of TiO$_{2}$ dispersion, we simulated the final optimized metalens and found its efficiency while considering the wavelength dependence of TiO$_{2}$ (see Fig. S17).  These efficiency values are shown in Fig. S10d and Fig. S11d. The efficiency values close to the maximum and minimum wavelength range are affected by dispersion, but the overall performance is not significantly affected. The average focusing efficiency values over the wavelength range when ignoring and considering the dispersion are 40.25\% and 38.98\%, respectively. Both these values are slightly larger than the ones reported in Ref.~\cite{Chung2020highNA}, showing the described method results in devices whose performance are comparable with state-of-the-art topology optimization.
\clearpage
\maketitle
\renewcommand{\theequation}{S\arabic{equation}}
\setcounter{equation}{0}

\section*{Supplementary Note 5: Gradient accuracy study}

\begin{figure*}
	\centering
	\includegraphics{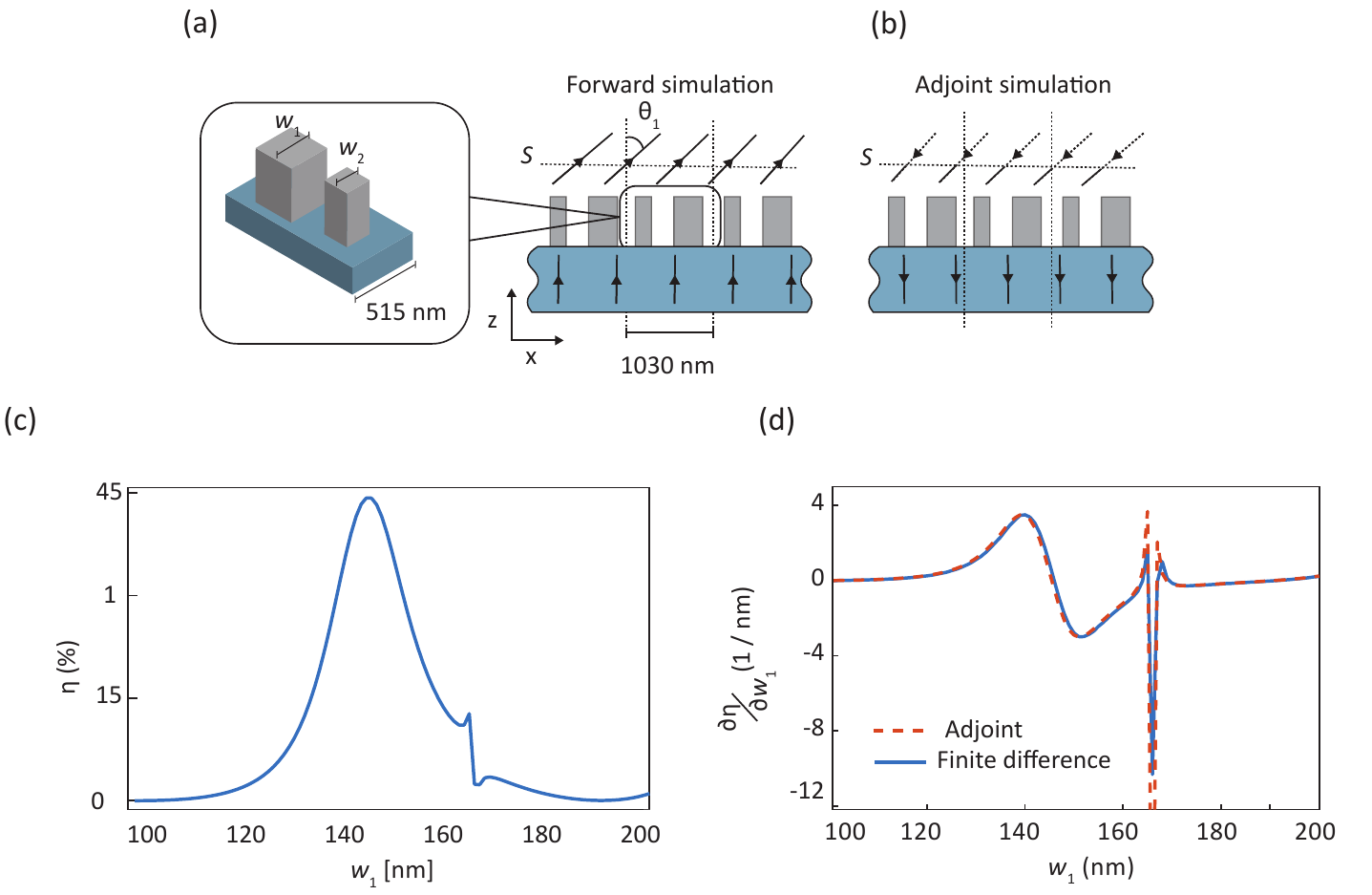}
	\caption{ Adjoint derivative accuracy validation.
		\textbf{(a)} Schematic of a grating’s forward simulation, and \textbf{(b)} its adjoint simulation.
		\textbf{(c)} Diffraction efficiency of the +1 order as a function of ${w}_1$. \textbf{(d)} Comparison of derivative values calculated by finite-difference and adjoint methods.}
	\label{two_post}
\end{figure*}

The convergence rate of the gradient ascent technique depends on the accuracy of the gradient calculation. To validate the gradient determined by our adjoint method, we compared the derivatives determined by Eq. (2) to a finite difference approximation of the simulated objective function. For this study we used a simple grating, shown in Fig.~S\ref{two_post}. The grating is a periodic metasurface with a $2\times1$ unit-cell composed of 560–nm-tall rectangular amorphous silicon (aSi) meta-atoms resting on a fused silica substrate as shown in the inset of Fig.~S\ref{two_post}a . The simulation domain is periodic in $x$ and $y$ directions.
In the $z$ direction, the structure is terminated above the output plane and below the input plane by perfectly matched layers (PML) with a thickness of 0.425~$\upmu$m. The design vector for this system consists of the widths of the meta-atoms: $\mathbf{p}=(w_1,w_2)$.

Simulations to verify the accuracy of the gradient vector estimated by the adjoint method were performed using a commercial finite element solver (COMSOL Multiphysics).
To find the necessary electric  fields in the forward simulation, the structure was excited by a surface electric current located inside the substrate.
Using conformal meshing on the boundaries of meta-atoms the exact boundary fields required to calculate the derivatives in forward and adjoint simulations are obtained with minimal discretization errors.

In the forward simulation, the electric current source generates a plane wave excitation hitting the structure from below.
In the adjoint problem, the structure was excited by a surface electric current placed on the output plane indicated in Fig.~S\ref{two_post} that
generates a plane wave hitting the structure exactly opposite to the direction of the first diffraction order.

We selected the desired field distribution $\mathbf{E}^\text{d}= \frac{1}{A} \exp{\left(-j\frac{2\pi}{\lambda_0}\sin{(\theta_1)}x\right)}\hat{x}$, where $A$ is the area of the grating unit-cell, $\theta_1=\sin^{-1}{\frac{\lambda_0}{2a}}={55.61}^\circ$ is the diffraction angle of the +1 diffraction order, $a=515$ nm the metasurface lattice constant, and $\lambda_0=850$ nm the free-space wavelength of the excitation. The complex projection $F$ is defined as in Eq. (1), where the integral was computed over a grating period, and the objective function was defined as $I=\left|F\right|^2$. In the forward simulation, the structure is excited by a normally-incident $x$-polarized plane wave as depicted in Fig.~S\ref{two_post}a.
The first diffraction order efficiency is calculated by $\eta=\frac{I \cos\left(\theta_1\right)}{2 Z_0 P_0}$,  where $I$ is the objective function, $P_0$ is the power of the incident field to the unit-cell, $Z_0$ is the impedance of free-space.

In the adjoint simulation (Fig.~S\ref{two_post}b), the structure is excited by a surface current on a plane $S$ normal to the $z$ direction that lies 90 nm above the top surface of the meta-atoms. The adjoint source generates a plane wave that arrives at the metasurface with incidence angle $\theta_1$.

Our choice of $\mathbf{E}^\text{d}$ makes $I$ proportional to the diffraction efficiency $\eta$ of the +1 diffraction order. Holding $w_2$ fixed at 100 nm, forward simulations were run varying $w_1$ between 100 nm and 200 nm in 1 nm increments. The diffraction efficiency as a function of $w_1$ is shown in Fig.~S\ref{two_post}c; its estimated derivatives, determined by both finite-difference and adjoint methods, are shown in Fig.~S\ref{two_post}d.  The good agreement between the results of adjoint and finite-difference methods confirms the accuracy of the adjoint method.

\bibliography{achemso-demo}
\bibliographystyle{ieeetr}

\clearpage
\section*{Supplementary Figures}

\clearpage

\clearpage

\begin{figure}
  \centering
  \includegraphics[width=3.5in]{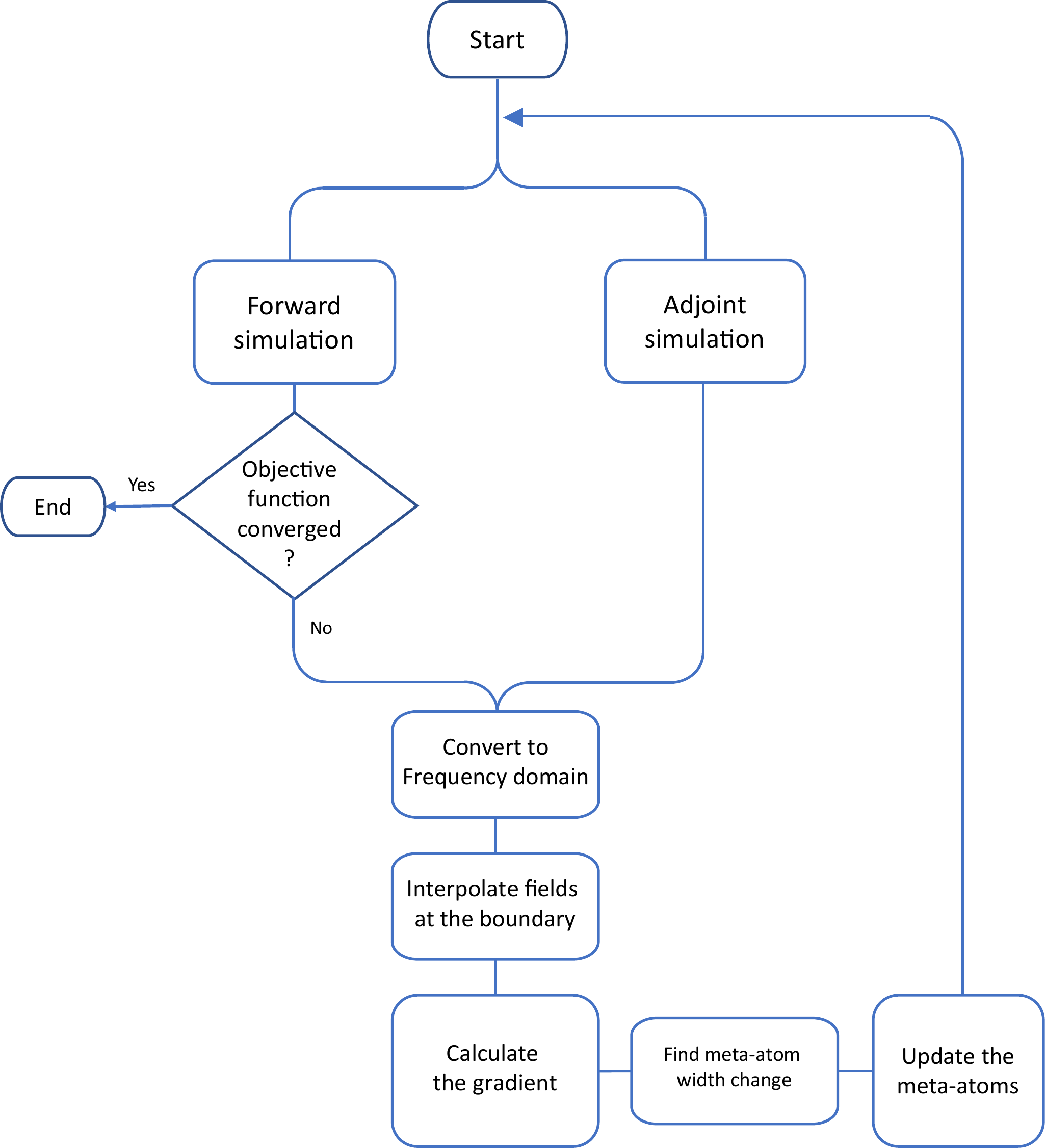}
  \caption{
    Flowchart of adjoint optimization.
    In each iteration, two simulations are done and the derivatives are calculated.
    The cycle is repeated until the design converged.
  }
  \label{Flow_chart}
\end{figure}
\clearpage

\begin{figure}
  \centering
  \includegraphics{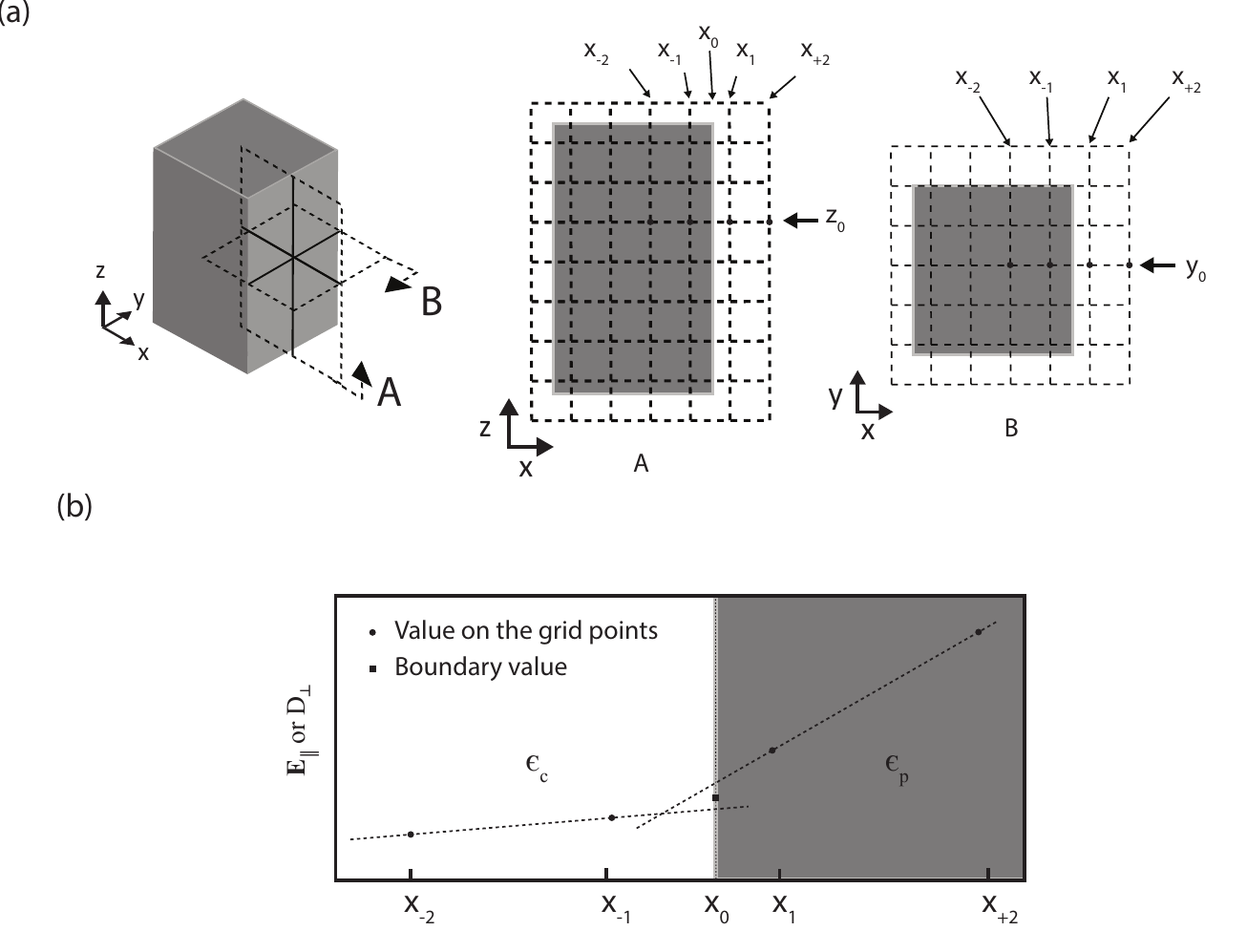}
  \caption{
    Interpolation scheme.
    (a) Yee grid and structural boundary.
    (b) A line is fit to each pair of grid points inside and outside the meta-atom.
    Then, the two lines are extrapolated and averaged to estimate the value on the boundary.
    In the averaging, the value estimated by each line has a weight equal to the inverse distance between the boundary and closest point on the line.
  }
  \label{interpolation}
\end{figure}
\clearpage

\begin{figure}
  \centering
  \includegraphics{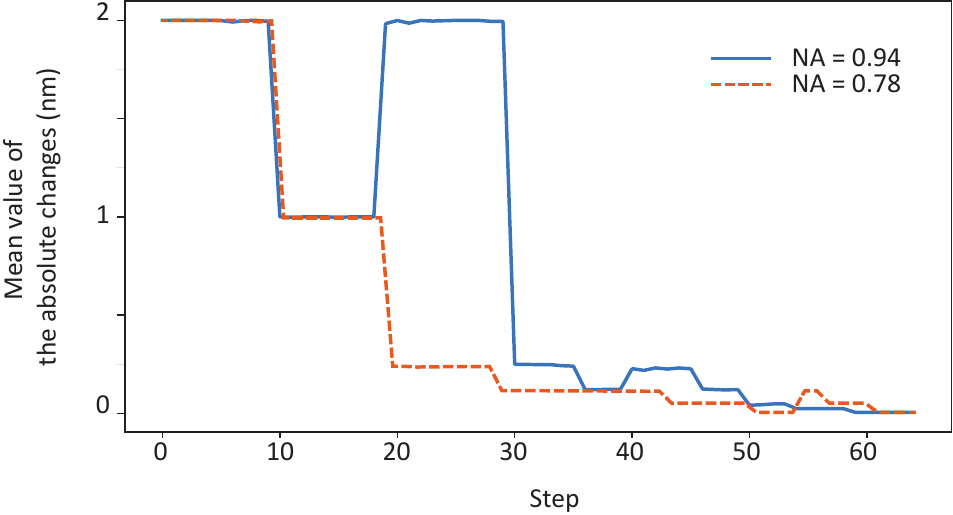}
  \caption{Average change in each iteration.
  }
  \label{step_size}
\end{figure}
\clearpage

\begin{figure}
  \centering
  \includegraphics{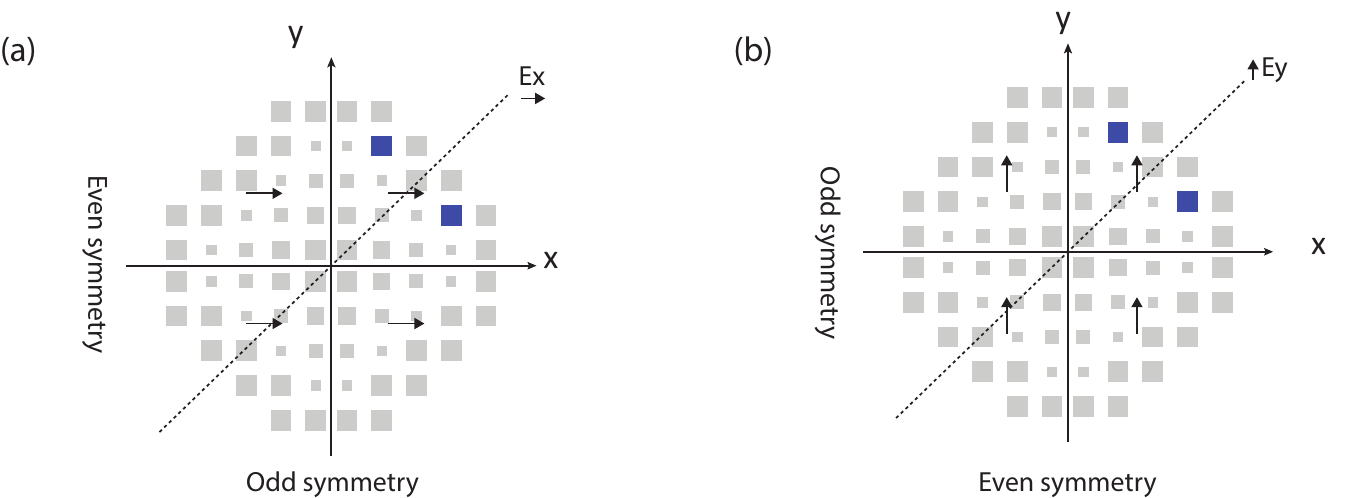}
  \caption{
    Symmetry planes in the simulations for each polarization.
    \textbf{(a)} Symmetries for $x$-polarized incident light.
    \textbf{(b)} The symmetry for $y$-polarized incident light.
    The geometry has a mirror symmetry for both axes and symmetry along the $x=y$ line.
    This symmetry is enforced when updating the structure.}
  \label{deriv_symmetry}
\end{figure}
\clearpage

\begin{figure}
  \centering
  \includegraphics[width=5in]{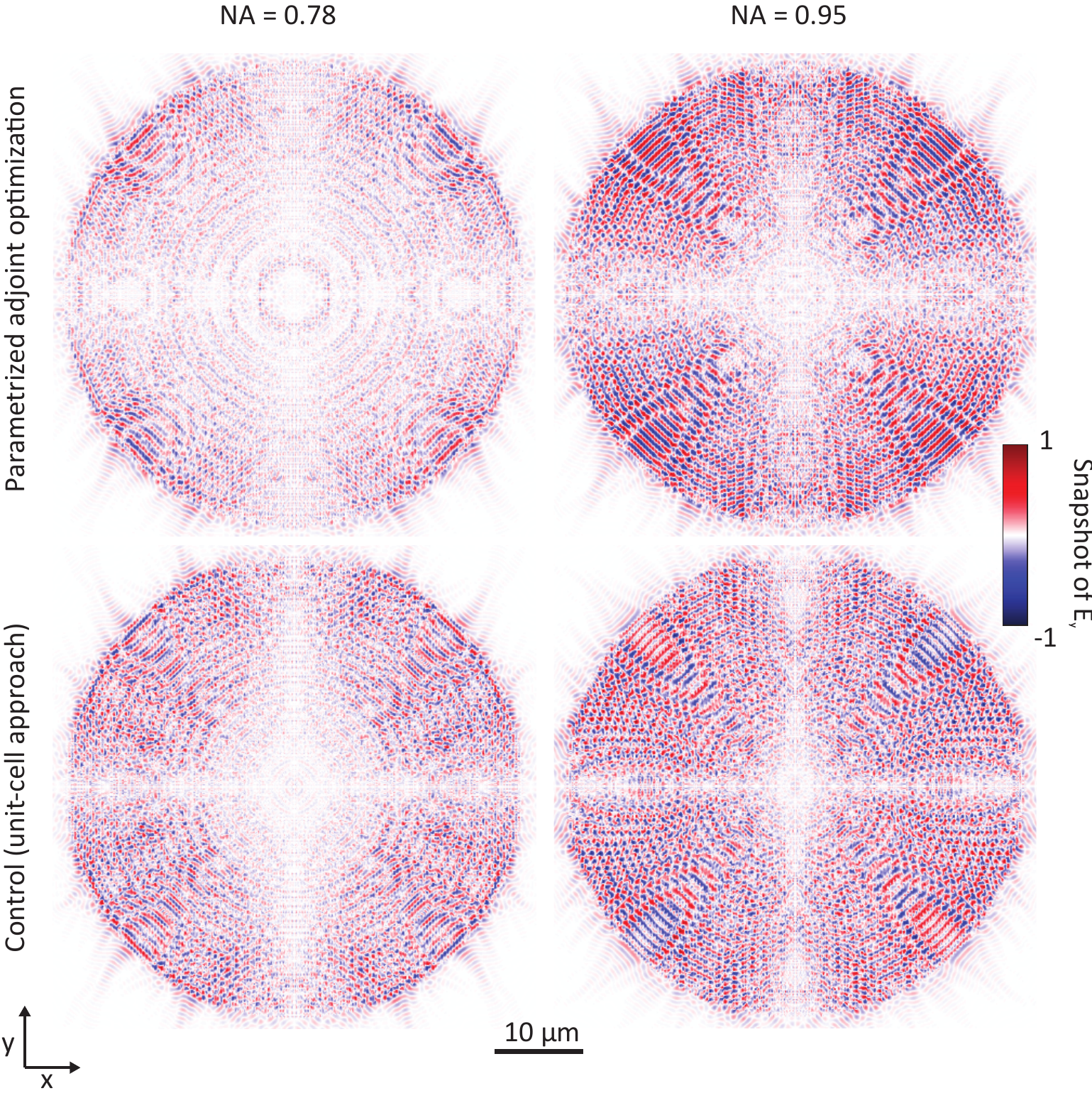}
  \caption{Cross-polarization components of the electric field $E_y$. The metalenses are excited with an $x$-polarized plane wave.}
  \label{cross_pol}
\end{figure}
\clearpage

\begin{figure}
	\centering
	\includegraphics{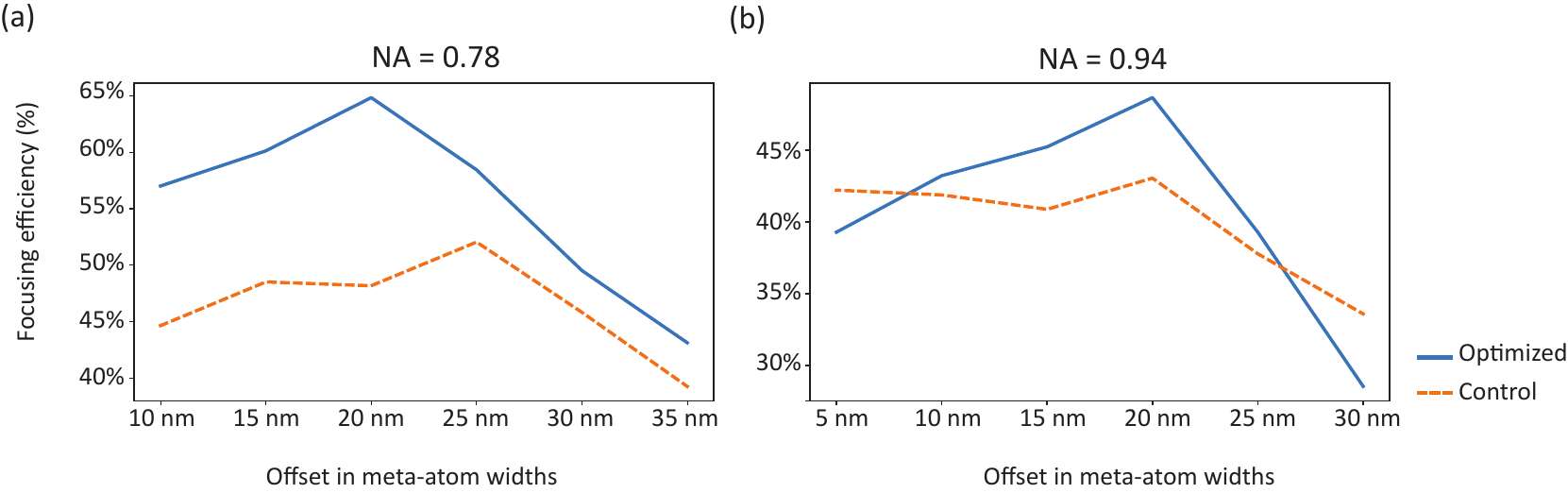}
	\caption{Measured efficiency for different values of  bias in width of the meta-atoms.  \textbf{(a)} for the metalens with NA=0.78, and \textbf{(b)} for NA=0.94.
		Adjoint optimized metalens is shown with solid line and control design is indicated by dashed line.
	}
	\label{bias_study}
\end{figure}
\clearpage

\clearpage

\clearpage

\clearpage

\begin{figure*}
	\centering
	\includegraphics[width=5in]{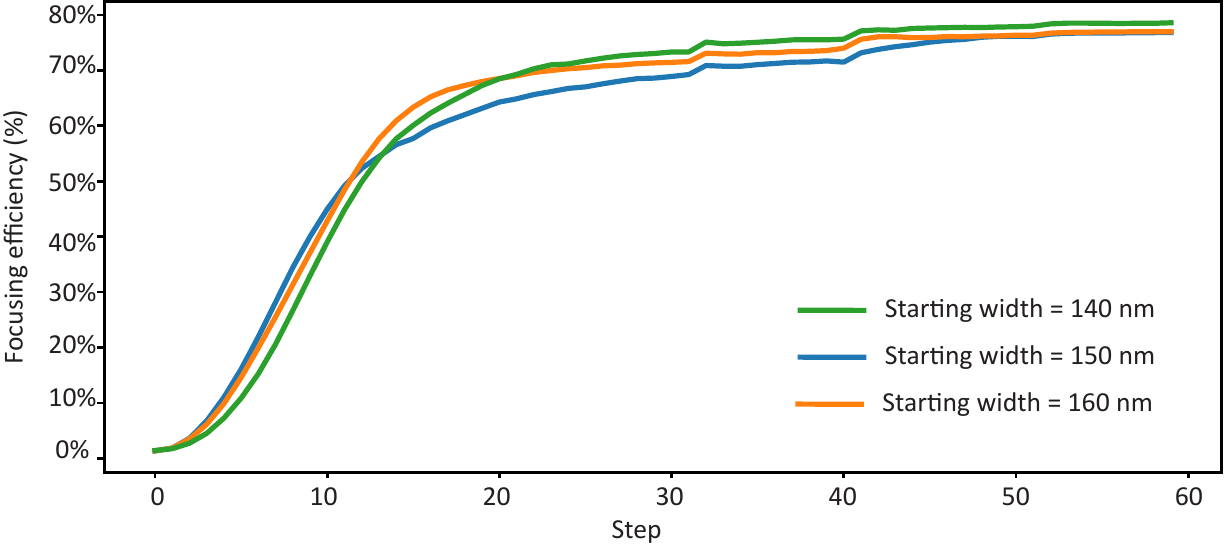}
	\caption{ Efficiency of metalenses with NA=0.78 and diameter of 50$\upmu$m at each step of the optimization. Each optimization started with different starting metasurface composed of a uniform array of meta-atom with equal width.}
	\label{starting_points}
\end{figure*}
\clearpage

\begin{figure}
  \centering
  \includegraphics[width=6in]{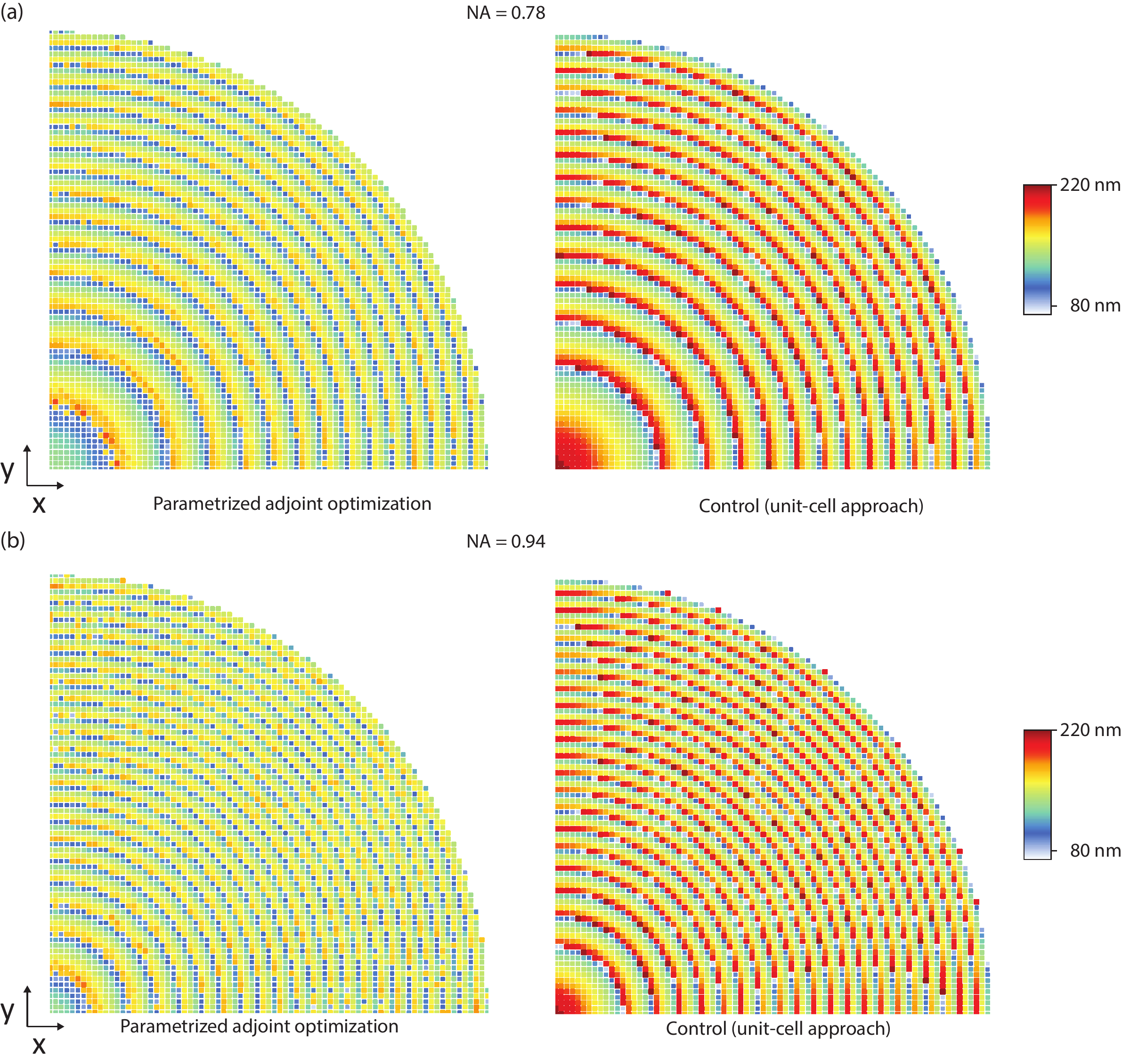}
  \caption{
    Color-coded plots of the meta-atom widths of the optimized and control metalens.
    \textbf{(a)} For NA of 0.78 and \textbf{(b)} for NA of 0.94.
    Because the structure is symmetric with respect to $x$ and $y$ axes, only the meta-atom widths in the first quadrant are shown.
  }
  \label{Quarters}
\end{figure}
\clearpage

\begin{figure}
  \centering
  \includegraphics{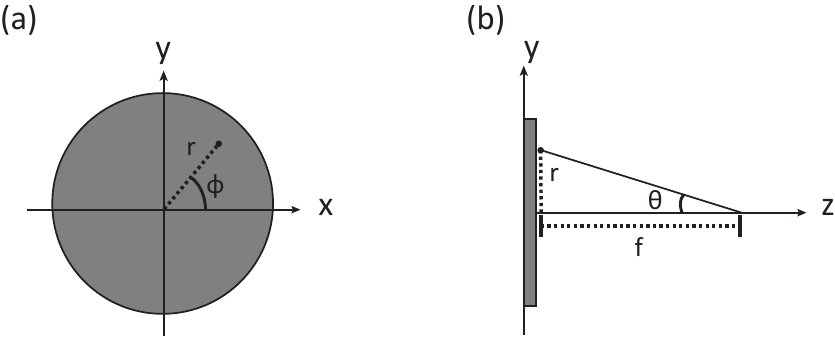}
  \caption{
    Relative coordinates for the desired output function in \textbf{(a)} for $xy$ plane and \textbf{(b)} in $yz$ plane.
  }
  \label{Cooridinates}
\end{figure}
\clearpage

\begin{figure}
  \centering
  \includegraphics[width=6.5in]{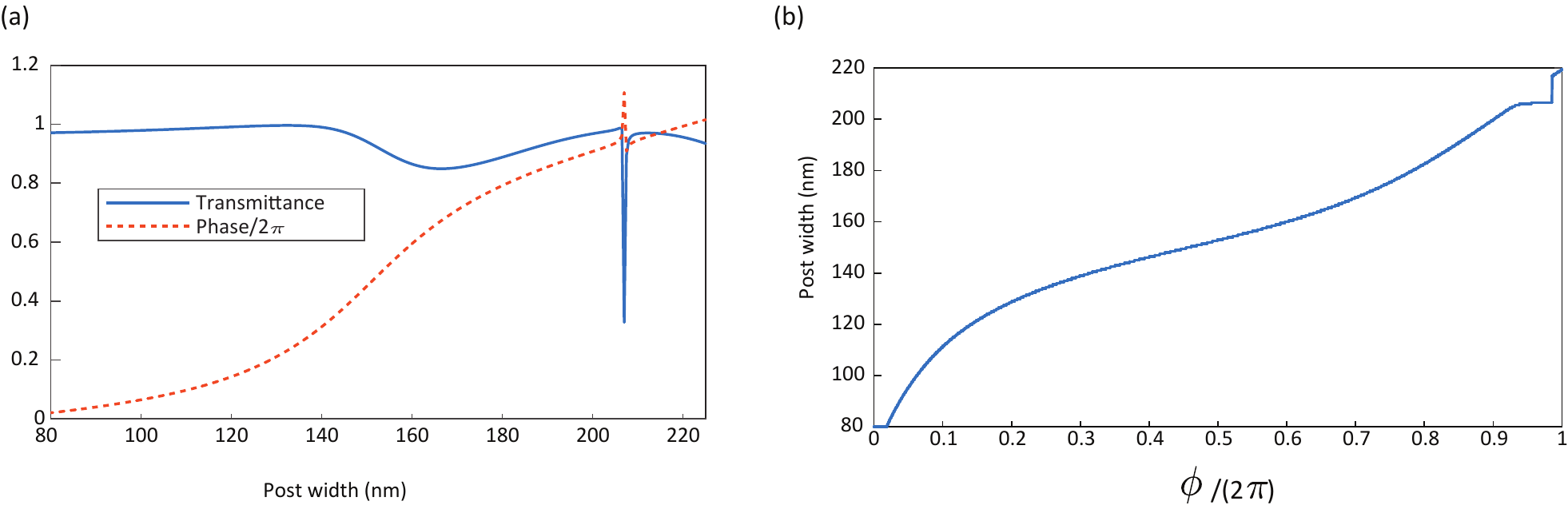}
  \caption{
    Conventional unit-cell approach design maps.
    (a) Simulated transmittance and phase of the transmission coefficient for control metalenses.
    (b) Design map used for the control metalenses.
  }
  \label{Conventional}
\end{figure}
\clearpage

\begin{figure}
  \centering
  \includegraphics[width=5in]{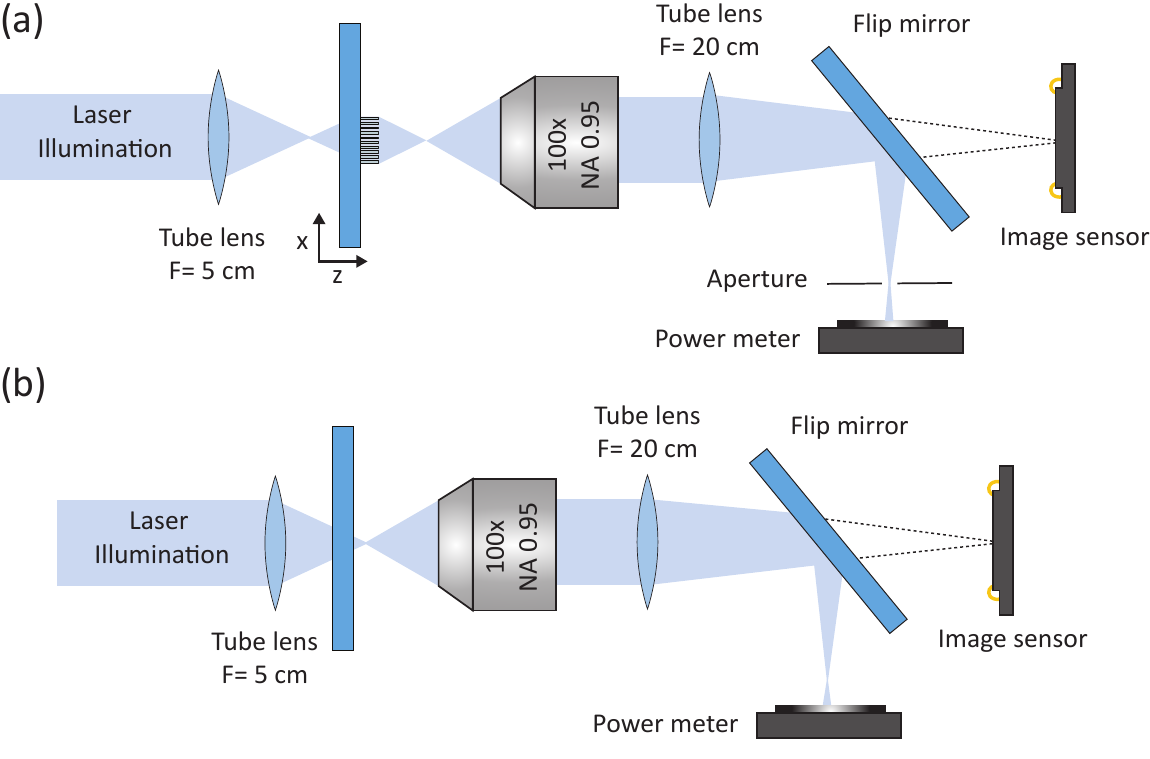}
  \caption{
    \textbf{(a)} Schematic drawing of the characterization setup.
    The fabricated metasurfaces were illuminated by a partially focused beam. The light passes through the sample and then is re-imaged on an image sensor using an objective lens (NA=0.95) and a tube lens.
    To measure the efficiency, a flip mirror was used to redirect the light toward an aperture with a diameter equal to the 7~$\upmu$m in the focal plane. \textbf{(b)} The setup is used to measure the incident power on the metalens.}
  \label{setup}
\end{figure}
\clearpage

\begin{figure}
	\centering
	\includegraphics{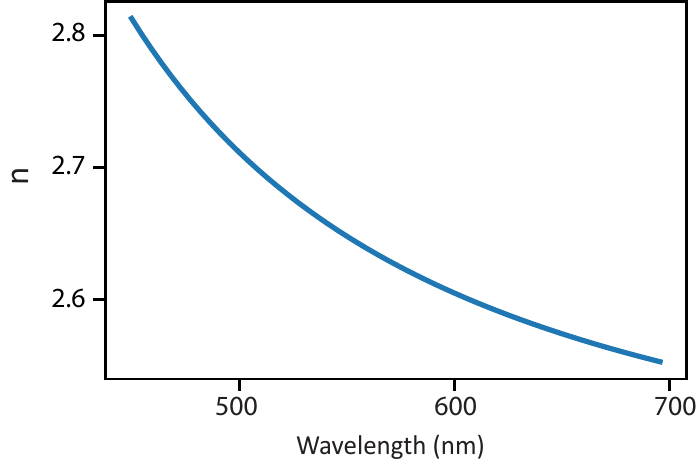}
	\caption{
		Refractive index of titanium dioxide.
		}
	\label{setup}
\end{figure}
\clearpage